\documentclass[%
reprint,twocolumn,
superscriptaddress,
amsmath,amssymb,
aps,
prstab,
]{revtex4-2}

\usepackage{amsmath,amssymb}
\usepackage{hyperref}
\usepackage{graphicx}
\usepackage{dcolumn}
\usepackage{bm}
\usepackage{gensymb}
\usepackage{amsmath}

\begin{document}
	
	
	
	\title{Numerical study on femtosecond electro-optical spatial decoding of transition radiation from laser wakefield accelerated electron bunches}

	\affiliation{Kansai Institute for Photon Science (KPSI), National Institutes for Quantum Science and Technology (QST), 8-1-7 Umemidai, Kizugawa-city, Kyoto, 619-0215, Japan}
	\affiliation{SANKEN, Osaka University, 8-1 Mihogaoka, Ibaraki, Osaka, 567-0047, Japan}
	\affiliation{Laser Accelerator R\&D, Innovative Light Sources Division, RIKEN SPring-8 Center, 1-1-1 Kouto, Sayo, Hyogo, 679-5148, Japan}

	\author{K. Huang}
	\email[Email: ]{huang.kai@qst.go.jp}
	\affiliation{Kansai Institute for Photon Science (KPSI), National Institutes for Quantum Science and Technology (QST), 8-1-7 Umemidai, Kizugawa-city, Kyoto, 619-0215, Japan}
	\affiliation{Laser Accelerator R\&D, Innovative Light Sources Division, RIKEN SPring-8 Center, 1-1-1 Kouto, Sayo, Hyogo, 679-5148, Japan}
	
	\author{Z. Jin}
	\affiliation{SANKEN, Osaka University, 8-1 Mihogaoka, Ibaraki, Osaka, 567-0047, Japan}
	\affiliation{Laser Accelerator R\&D, Innovative Light Sources Division, RIKEN SPring-8 Center, 1-1-1 Kouto, Sayo, Hyogo, 679-5148, Japan}

	\author{N. Nakanii}
	\affiliation{Kansai Institute for Photon Science (KPSI), National Institutes for Quantum Science and Technology (QST), 8-1-7 Umemidai, Kizugawa-city, Kyoto, 619-0215, Japan}
	\affiliation{Laser Accelerator R\&D, Innovative Light Sources Division, RIKEN SPring-8 Center, 1-1-1 Kouto, Sayo, Hyogo, 679-5148, Japan}
	
	\author{T. Hosokai}
	\affiliation{SANKEN, Osaka University, 8-1 Mihogaoka, Ibaraki, Osaka, 567-0047, Japan}
	\affiliation{Laser Accelerator R\&D, Innovative Light Sources Division, RIKEN SPring-8 Center, 1-1-1 Kouto, Sayo, Hyogo, 679-5148, Japan}

	\author{M. Kando}
	\affiliation{Kansai Institute for Photon Science (KPSI), National Institutes for Quantum Science and Technology (QST), 8-1-7 Umemidai, Kizugawa-city, Kyoto, 619-0215, Japan}
	\affiliation{Laser Accelerator R\&D, Innovative Light Sources Division, RIKEN SPring-8 Center, 1-1-1 Kouto, Sayo, Hyogo, 679-5148, Japan}

	\date{\today}

	\begin{abstract}
		This numerical study is focused on electro-optic (EO) spatial decoding of transition radiation (TR) produced by a relativistic electron bunch passing through a metal foil. The calculations included the imaging of polychromatic transition radiation from an electron bunch and the process of EO spatial decoding. From an experimental perspective, a careful examination of the calculation approach of the data analysis is essential. Therefore, to thoroughly understand the process of signal generation and examine the possibility of adopting a less time-consuming treatment, comparative studies were conducted on detailed and simplified models of both transition radiation imaging and EO signal generation. All calculations are defined in SI units for the convenience of experimental measurements. For TR imaging, the results suggest that the simplified analytical model is sufficient to perform polychromatic calculations with considerable accuracy. For EO spatial decoding, we discussed the process of EO signal generation using 1D and 2D models. We found that the 1D model was sufficient for rapid data analysis. Furthermore, the temporal energy chirp was demonstrated to have a minimal impact on the shape of the EO signal. Because both the transverse and longitudinal profiles can be calculated with arbitrary distributions, this numerical study can facilitate measurements of 3D electron charge density profiles in both laser wakefield acceleration and conventional accelerator research. 
	\end{abstract}
	
	
	\maketitle
	
	\section{Introduction}
	Laser wakefield acceleration\cite{tajima1979laser,geddes2004high,faure2004laser,mangles2004monoenergetic}, with an inherent ultrashort timescale and ultrahigh acceleration gradient, has attracted great interest worldwide. With the incidence of a high-power femtosecond (fs) laser beam on an underdense plasma, electron plasma waves with a wavelength of a few tens of micrometers ($\mu$m) and an electric field strength of $>$ 100 GeV/m can be stimulated. Trapped electrons achieve energies of over GeV \cite{leemans2006gev,leemans2014multi,gonsalves2019petawatt,wang2013quasi,kim2013enhancement,mirzaie2015demonstration} within a centimeter. In 2021, free-electron lasing at 27 nm driven by an laser wakefield accelerator (LWFA) was demonstrated via the generation of high-quality electron bunches \cite{wang2016high,wang2021free}. \par
	
	The three-dimensional (3D) charge density distribution of a relativistic electron bunch is an important parameter for secondary radiation sources of an LWFA, such as betatron X-rays \cite{albert2014laser,huang2016resonantly}, inverse Compton scattering \cite{yan2017high}, and X-ray free electron lasers (XFELs) \cite{nakajima2008compact}. The duration of the electron bunch partially determines the temporal resolution of a pump-probe experiment. The gain length of lasing processes in an undulator \cite{ZhirongHuang} is affected by the 3D charge density.  \par
	
	Electro-optic (EO) sampling \cite{shan2000single} \cite{yan2000subpicosecond,wilke2002single,berden2004electro,berden2007benchmarking,steffen2009electro,cavalieri2005clocking,yang2009electron,wang2017temporal,scoby2010electro,jamison2006electro,steffen2007electro,casalbuoni2008numerical} was applied to accelerator studies 20 years ago owing to its single-shot and non-destructive capabilities. Pioneering numerical studies focused on this convenient method can be found in \cite{steffen2007electro,casalbuoni2008numerical}. When the electric field carrying the temporal information of the electron bunch propagates through an EO crystal, the effective principal axis of the crystal rotates because of the Pockels effect. With the incidence of a probe laser beam on a crystal, time information of the electron bunch can be reconstructed by analyzing the phase retardation. The longitudinal distribution of electrons is encoded transversely into the probe laser beam by setting a relative angle between the probe laser and the signal field. This is referred to as the “EO spatial decoding technique” \cite{wang2017temporal}.  \par
	
	EO sampling of transition radiation (TR) of an electron bunch has been conducted to measure the relative electron longitudinal profile \cite{JvanTilborg,ADebus} or the field strength of a terahertz pulse \cite{CTR-EOtsinghua}. However, simultaneous measurement of the absolute current and transverse profiles has not been attempted. By the incidence of the TR pulse onto the EO crystal and the performing the spatial decoding, the TR field produced by the electron bunch is measured both temporally and spatially. We named the detection method as the ``TR-EO" method in this study.\par
	
	In this article, we performed numerical calculations of EO spatial decoding of TR produced by a relativistic electron bunch passing through a metal foil. The calculations involved the imaging of polychromatic TR from electron bunches with arbitrary shapes and energies and the EO spatial decoding process, including TR absorption by the crystal, smearing, and phase mismatch, which are affected by the relative angle and speed difference between the TR and the probe laser. All calculations in this study are defined in SI units for ease of implementation. Detailed and simplified models were discussed for the calculations of both the TR imaging and EO signal generation. Through the comparison of results, the feasibility of the simplified models for a quicker analysis was demonstrated. In addition, we calculated the impact of temporal energy chirps of electrons on the EO signals and concluded that the current profiles dominated the shapes of the EO signals. \par
	
	This paper is organized as follows: the experimental setup of this numerical study is described in Sec. II. The calculation of the polychromatic TR field via a detailed Huygens--Fresnel diffraction calculation and a simplified analytic model based on Fraunhofer approximation is described in Sec. III. The EO spatial decoding process is described in Sec. IV using a simplified model and a detailed 2D interpolation method. The impact of the temporal energy chirp of electrons on the EO signal is discussed in Sec. IV. Thereafter, noise in transition radiation generated by the optical system is calculated in Sec. V. Finally, the discussion and conclusions of this study are presented in Sec. VI. \par
	
	\section{Experimental concept of this numerical study}
	
	\begin{figure}[ht] 
		\centering
		\includegraphics[width=7.5 cm]{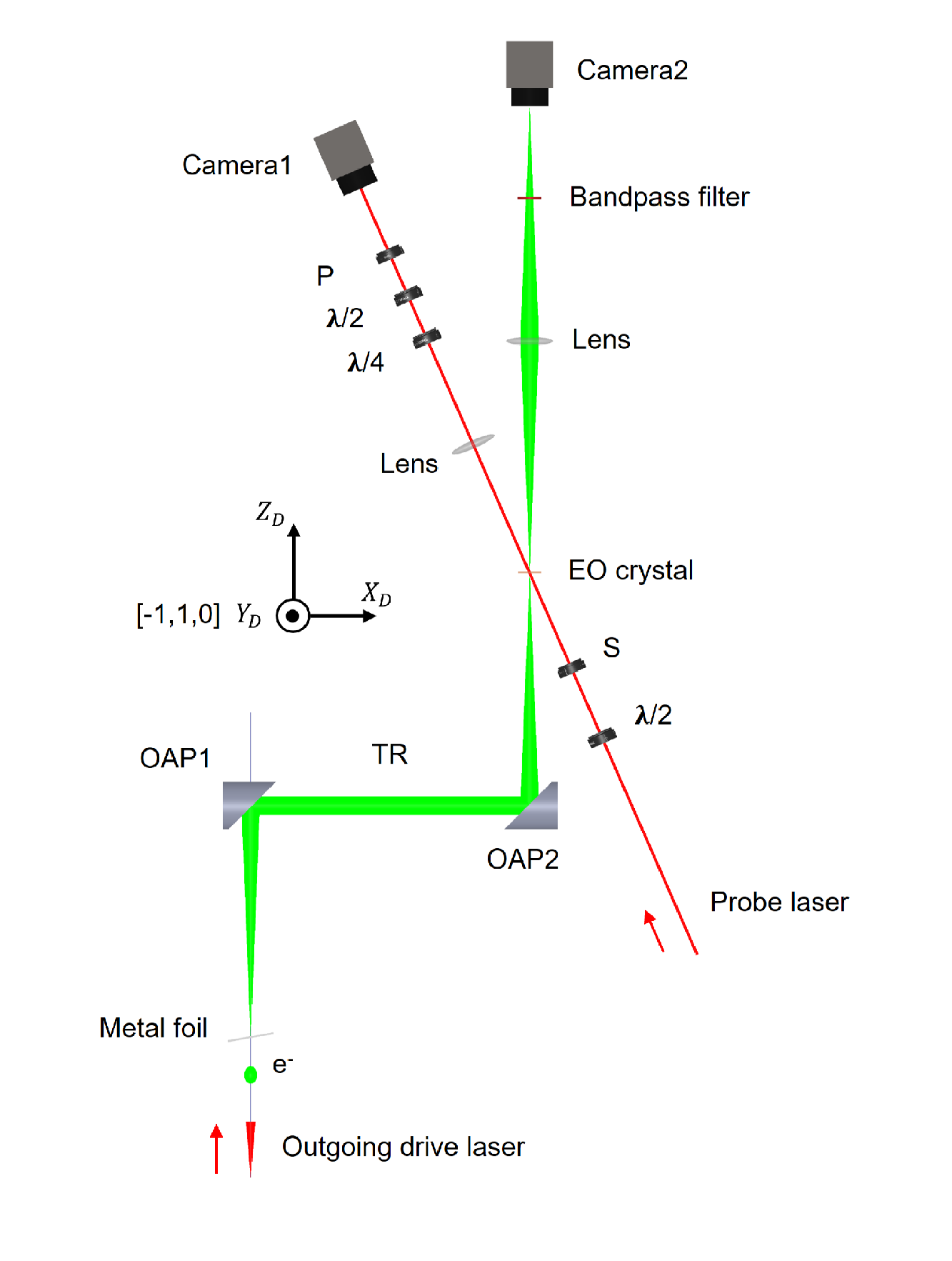}
		\caption{Conceptual scenario of the TR-EO spatial decoding set-up for an LWFA. e$^{-}$: electron bunch; OAP: Off-axis-parabolic gold mirror; S: polarizer before EO crystal; P: polarizer after EO crystal; $\lambda$/2: half waveplate; $\lambda$/4: quarter waveplate; and BP: bandpass filter. The inset shows the coordinates in the imaging plate. The [-1,1,0] axis of the crystal is along the $Y_D$ direction. The polarization of the incident probe laser is also parallel to the $Y_D$ direction.}
		\label{f1}
	\end{figure} 
	
	Numerical calculations were conducted to quantitatively explain the aspects that can be achieved using the experimental concept illustrated in Fig. 1. In the LWFA, a drive laser beam was incident on a gas target to generate an electron beam. Therefore, electron bunches propagated with the outgoing drive laser beam. However, an intense drive laser can result in damage and noise during particle parameter measurements in the LWFA. Therefore, a metal foil (e.g., aluminum, or stainless steel) was placed in the electron beam path to eliminate noise generated by the drive laser beam and produce TR. Subsequently, the TR emission was imaged using two off-axis parabolic (OAP) mirrors and was incident onto an EO crystal. Typically, crystals with a zinc-blende structure are used. An ultrashort probe laser beam with a small incident angle $\theta_p$ to the EO crystal was used to perform spatial decoding. The [-1, 1, 0] axis of the crystal was parallel to the polarization direction of the probe laser and orthogonal to the plane formed by the probe laser beam and the TR propagation direction. Further, half-wave and quarter-wave plates were inserted to perform the near-cross-polarization \cite{steffen2007electro} detection. The TR passing through the crystal was then imaged again to measure the optical transition radiation (OTR). Because the TR field had a wavelength range from visible to infrared, it is recommended that the EO crystal be placed inside a vacuum chamber to avoid unwanted absorption from the vacuum window and air. For the same reason, the gold-coated OAP was used instead of a lens to deliver the TR pulse beam. \par
	
	\begin{figure*}[ht]
		\centering
		\includegraphics[width=17 cm]{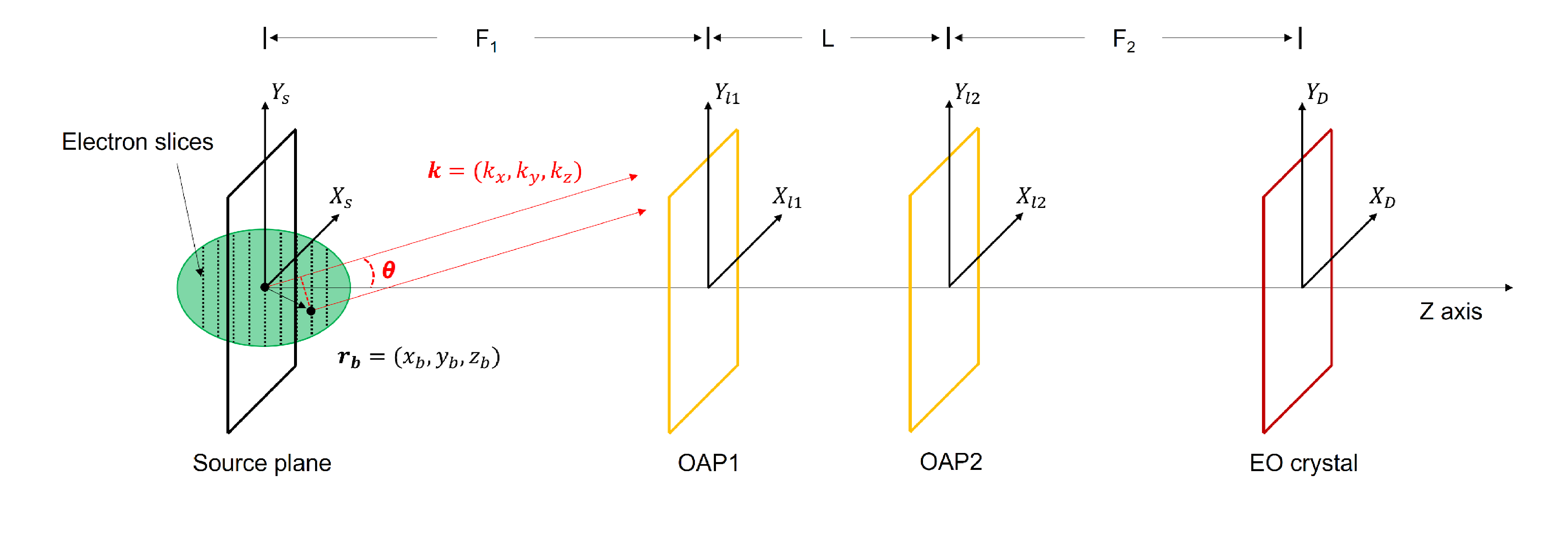}
		\caption{Calculation geometry of the imaging of TR of an electron bunch. OAP1 and OAP2 have effective focal lengths of $F_1$ and $F_2$, respectively. The distance between the OAPs is $L$.}
		\label{f2}
	\end{figure*}
	
	\section{Three-dimensional TR electric field in the coordinates ($X_D$, $Y_D$, $T$)}
	In this section, we present a 3D calculation of the TR field imaged on the EO crystal. The calculation geometry is shown in Fig. 2. An electron bunch passed through a radiator (source plane), and TR was produced. The produced TR was collected by the first OAP (OAP1) with a focal length of $F_1$. The source point was placed at the focus of OAP1. Consequently, the collimated TR light was focused by OAP2 on the EO crystal. Here, ($X_s,Y_s$) and ($X_D, Y_D$) denote the mesh grids in the source and detector planes (EO crystals), respectively, and ($X_{l1}, Y_{l1}$) and ($X_{l2}, Y_{l2}$) are the mesh grids in the planes of OAP1 and OAP2, respectively, with the OAPs considered as thin lenses. The mesh grids were arranged symmetrically with the centers set at (0, 0). The distance between OAP1 and OAP2 was denoted as $L$. Further, $\bm{r_b}=(X_b,Y_b,Z_b)$ denotes the relative position of a single electron inside the electron bunch. The wave number of the TR was $k=\sqrt{k_x^2+k_y^2+k_z^2}$ and it had a relative angle of $\theta$ relative to the z-axis. The direction of the [-1,1,0] axis of the crystal was along the ``$Y$" axes in Fig. 2. For EO spatial decoding, the polarization of the probe laser beam was set along the [-1,1,0] axis of the EO crystal to achieve maximum phase retardation. \par
	
	In this study, the third dimension ``$T$" or ``$\omega$" was assumed to be decoupled from the transverse beam profile. First, we calculated the 2D spatial distribution of the TR field $E_y (X_D, Y_D, \omega)$ for a single-frequency component. Subsequently, the 2D image was multiplied by the longitudinal form factor $F_z(\omega)$. Finally, we performed a Fourier transform of $E_y (X_D, Y_D, \omega)$ in the $\omega$ dimension to obtain $E_y (X_D, Y_D, T)$ distribution. \par
	
	By treating the metal foil as an ideal conductor, the TR of a single electron can be calculated as the total reflection of the self-field (superposition of pseudo-photons). Here, without loss of generality, when passing through the foil, a single electron at position (0, 0) of the source plane is considered. The self-field expression (further details presented in Appendix A) of an electron in the frequency and wavenumber domains is expressed as, 
	
	\begin{widetext}
		\begin{subequations} 		
			\begin{equation}\label{Erwxys1}
				E^s(X_s,Y_s)=i\dfrac{e}{(2\pi)^2\epsilon_0 v}\iint{dk_xdk_y}\exp[i(k_xX_s+k_yY_s)]\dfrac{(k_x,k_y)}{k_x^2+k_y^2+\alpha^2}
			\end{equation}
			\begin{equation} \label{Erwxys2}
				E^s(X_s,Y_s)=-\dfrac{e\alpha}{2\pi \epsilon_0 v}\dfrac{(X_s,Y_s)}{\sqrt{X_s^2+Y_s^2}}K_1(\alpha\sqrt{X_s^2+Y_s^2})
			\end{equation}
		\end{subequations}
	\end{widetext} 
	\noindent
	
	where $\alpha=\omega/\gamma v$, where $v$ is the electron velocity, $K_1$ is a Bessel function of the second type, and $\epsilon_0$ denotes the electric constant. Eq.~(\ref{Erwxys1}) and Eq. (\ref{Erwxys2}) are equivalent and used as the source field of the TR from an electron. \par

	\subsection{2D spatial imaging of the TR field}
	Calculations related to OTR imaging have been investigated in accelerator research to measure the transverse size of electron bunches. Both incoherent and coherent OTR imaging \cite{xiang2007theoretical,CastellanoOTR,casalbuoni2009ultrabroadband,loos2008observation,potylitsyn2020image} have been focused upon. In these studies, a far-field model based on Fraunhofer assumption was widely used because the TR was specially treated in the optical range. However, because the EO sampling of the TR covers the wavelength range of zero to a few hundred THz, the validity of the Fraunhofer model should be discussed. \par
	
	\subsubsection{Detailed numerical diffraction calculation based on the Huygens--Fresnel principle}
	
	We calculated the 2D TR field $E_y (X_D, Y_D)$ based on the Huygens--Fresnel principle\cite{goodman1988introduction} as follows:
	\begin{equation}\label{HFresnel}
		U(x_2,y_2)=\frac{z}{i\lambda}\iint_{\Sigma_1}U(x_1,y_1)\frac{\exp(ikr_{12})}{r^2_{12}}dx_1dy_1
	\end{equation}
	\noindent
	where ``1" and ``2" are the indices of planes ``1" and ``2", respectively, with a distance of $z$, and $U$ is the scalar field distribution. Further, ($x_1$, $y_1$) and ($x_2$, $y_2$) are points on the two planes with the distance between them being $r_{12}=\sqrt{z^2+(x_2-x_1)^2+(y_2-y_1)^2}$. The reasons for choosing the Huygens--Fresnel principle instead of Fresnel approximation are as follows: (i) the para-axial condition might not be fulfilled if the electrons have low energy because the TR has an angular peak at $\theta_{peak}\sim 1/\gamma$; and (ii) the acceptance angle of OAP1 is occasionally large in the experiment. \par
	
	\begin{figure*}[ht]
		\centering
		\includegraphics[width=18 cm]{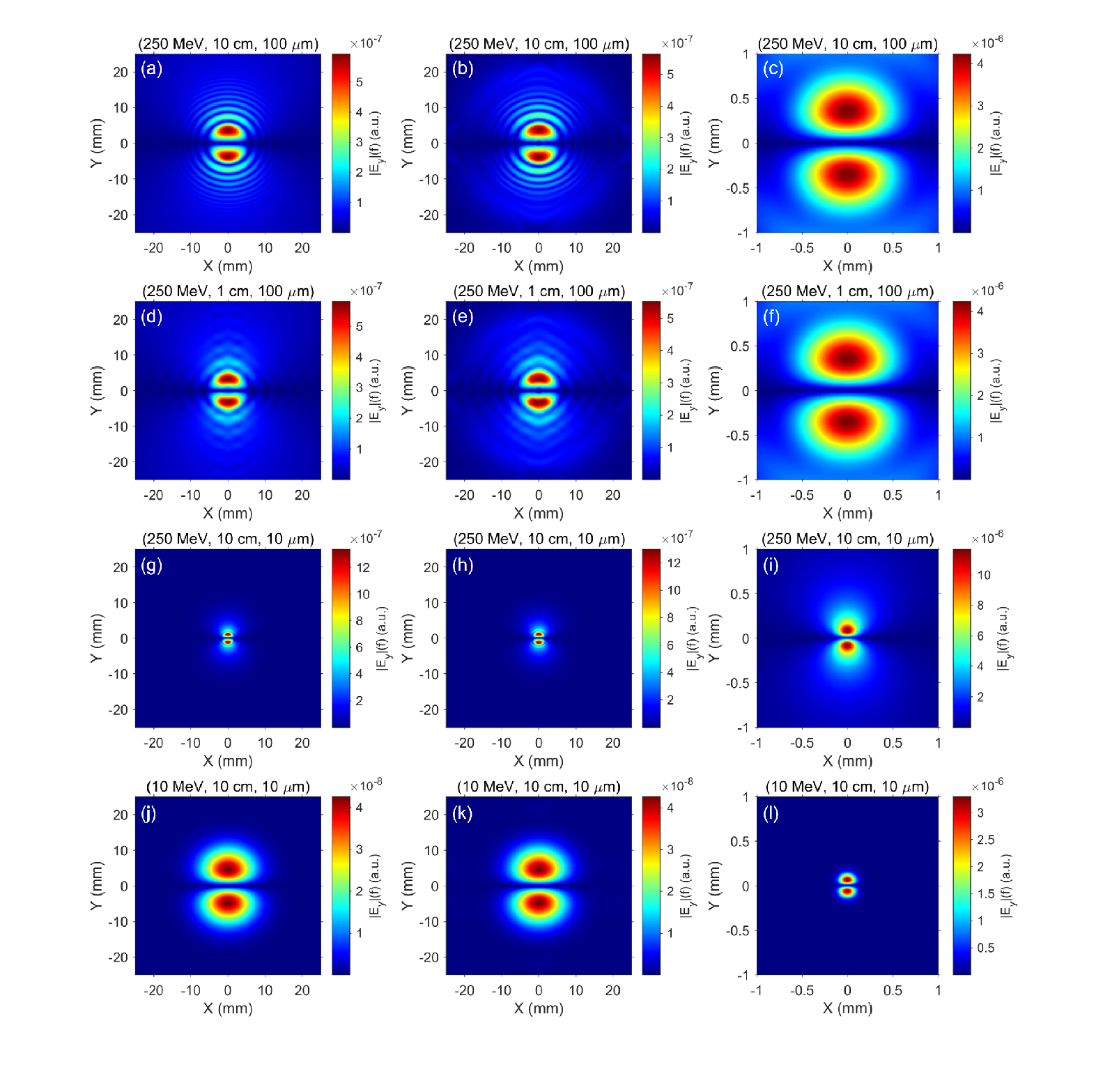}
		\caption{Calculation results based on the Huygens--Fresnel principle. The $|E_y (X,Y,f)|$ at the positions of OAP1 (a, d, g, and j), OAP2 (b, e, h, and k), and the EO crystal (c, f, i, and l) are plotted. The first row (a--c) shows the results of \{electron energy, metal foil size, wavelength\} = \{250 MeV, 10 cm, 100 $\mu$m\}.The second row (d--f) shows the results of \{250 MeV, 1 cm, 100 $\mu$m\}. The third row (g--i) shows the results of \{250 MeV, 10 cm, 10 $\mu$m\}. The fourth row (j--l) shows the results of \{10 MeV, 10 cm, 10 $\mu$m\}.} 
		\label{f3}
	\end{figure*}
	
	\begin{figure*}[ht]
		\centering
		\includegraphics[width=17 cm]{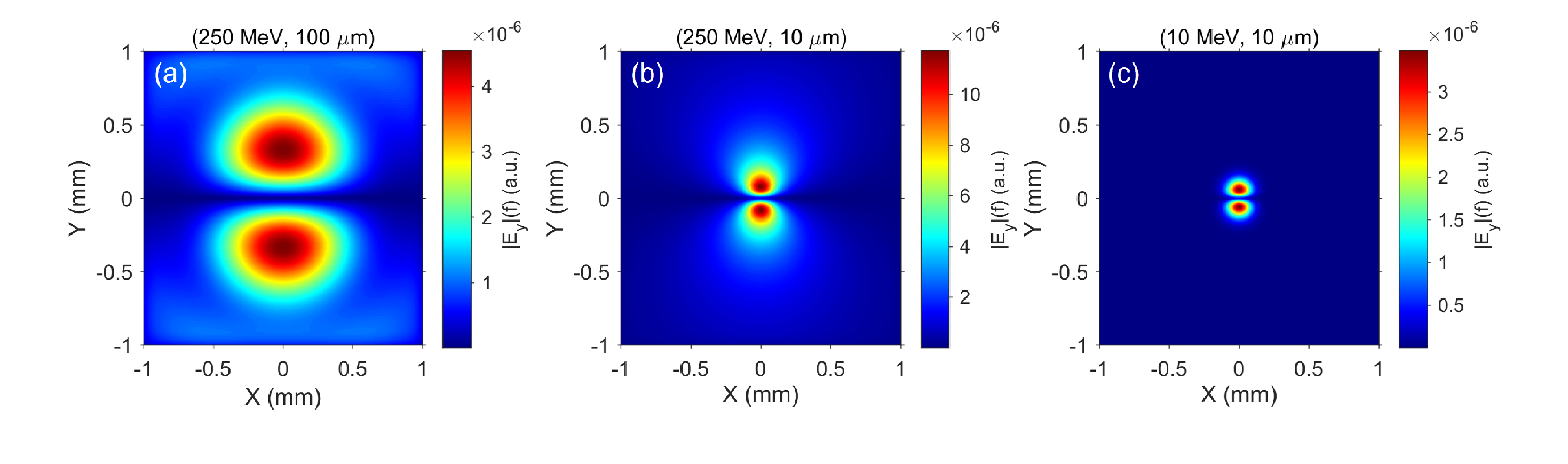}
		\caption{Calculation results of $|E_y (X,Y,f)|$ at the EO crystal based on an analytic model based on Fraunhofer assumption. Here, (a) shows the result of \{electron energy, wavelength\} = \{250 MeV, 100 $\mu$m\}, (b) shows the result of \{250 MeV, 10 $\mu$m\}, and (c) shows the result of \{10 MeV, 10 $\mu$m\}.} 
		\label{f4}
	\end{figure*}
	
	Equation~(\ref{HFresnel}) can be treated easily as a convolution. We defined the transfer function as:
	
	\begin{equation}\label{Hfunction}
		H(x,y)=\frac{z}{i\lambda}\frac{\exp(ik\sqrt{z^2+x^2+y^2})}{z^2+x^2+y^2}
	\end{equation}
	\noindent
	The Huygens--Fresnel principle expressed in Eq.(\ref{HFresnel}) can be rewritten as follows:
	\begin{equation}\label{HFresnelconv}
		U(x_2,y_2)=\iint_{\Sigma_1}U(x_1,y_1)H(x_2-x_1,y_2-y_1)dx_1dy_1
	\end{equation}
	\noindent
	By setting both planes with the same mesh grid, the diffraction can be calculated using the following Fourier transformation:
	\begin{equation}
		U_2=U_1\ast H=\mathcal{F}^{-1}[\mathcal{F}(U_1)\times\mathcal{F}(H)]
	\end{equation}
	\noindent
	where $\mathcal{F}$ and $\mathcal{F}^{-1}$ denote the Fast Fourier Transfomation (FFT) and Inverse FFT, respectively. Three transfer functions $\{H_1, H_L, H_2\}$ existed in the system with different longitudinal distances as $\{F_1, L, F_2\}$. \par

	There were three optical apertures with sharp edges (step functions) in the imaging system: (i) a finite size of the metal foil $D_s$, (ii) a finite size of OAP1; and (iii) a finite size of OAP2. In this study, we set up two OAPs with the same radius $R_l$ and focal length $F$. These apertures can be described by the following matrices:
	
	\begin{equation}
		\begin{aligned}
			P_s(X_s,Y_s)&=\{|X_s|, |Y_s|\}<D_s/2 \\
			P_l(X_l,Y_l)&=\sqrt{X_l^2+Y_l^2}<R_l
		\end{aligned}
	\end{equation}
	\noindent
	where $P^s(X_s,Y_s)$ is the square pupil (metal foils often have square shapes) at the source plane and $P^l(X_l,Y_l)$ is the round pupil created by the OAPs. \par
	
	An OAP with an effective focal length $F$ can be considered an ideal lens without spherical aberrations (explanations are provided in Appendix B). The converging effect of a lens results in a relationship between the field at the left side of the lens $E^{ll}$ and the right side of the lens $E^{lr}$ as follows:
	\begin{equation}
		E^{lr}(X_l,Y_l)=E^{ll}\exp(-ik\dfrac{X_l^2+Y_l^2}{2F})
	\end{equation}
	\noindent
	where $F$ is the focal length of the lens. The wavefront difference before and after the OAP is expressed as,
	\begin{equation}\label{lenstransform}
		PT_l=\exp(-ik\dfrac{X_l^2+Y_l^2}{2F})
	\end{equation}
	\noindent
	
	Considering the transverse distribution of $g_\perp(X_s,Y_s)$, the Coulomb field of one slice in the electron bunch at the source plane is a convolution, which is expressed as follows:
	\begin{equation}
		Cou^s (X_s, Y_s)=g_\perp\ast E^s
	\end{equation}
	\noindent
	Here, we used the expression of the field of a single electron $E^s$, as expressed in Eq. ~(\ref{Erwxys2}). The transverse distribution was arbitrary. The TR field on the left side of the OAP1 was calculated as follows:
	\begin{equation}
		E^{ll1}(X_{l1}, Y_{l1})=(P_s \times Cou^s)\ast H_1
	\end{equation}
	\noindent
	The diffractive propagation of the TR from the left side of OAP1 to the left side of OAP2, including the phase shift and aperture of OAP1, was calculated as follows:
	\begin{equation}
		E^{ll2}(X_{l2}, Y_{l2})=(P_l \times PT_l \times E^{ll1})\ast H_L
	\end{equation}
	\noindent
	Similarly, TR field propagation from the left side of OAP2 to the EO crystal was calculated as follows:
	\begin{equation}
		E^D(X_{D}, Y_{D})=(P_l \times PT_l \times E^{ll2})\ast H_2
	\end{equation}
	
	Equations (9--12) show the quantitative calculation procedures for the diffractive propagation of the TR field. We set $R_l$ = 25.4 mm and $F$ = 190  mm for the current and subsequent calculations. The distance between both the OAPs was set to $L$ = 40 cm. The TR field profiles of $E^{ll1}$, $E^{ll2}$, and $E^D$ were calculated to check the diffraction patterns, as shown in Fig. 3. We first performed calculations using an electron energy of 250 MeV with a TR wavelength of 100 $\mu$m. In this case, the effective electron radius\cite{casalbuoni2009ultrabroadband,xiang2007theoretical} was $\gamma\lambda\sim$ 5 cm. A transverse beam of size (50, 50) $\mu$m was assumed. The results of metal foil sizes of $D_S$ = \{1, 10\} cm were compared, as shown in Figs. 3(a--f). Although the diffraction patterns at the positions of the OAPs were slightly different, the TR field distributions in the EO crystal were almost identical. This calculation facilitated an understanding of the diffraction of the TR field at intermediate positions inside the optical system. Furthermore, this indicates that, in most cases, the variation in the metal foil size should merely result in a difference in the field strength at the imaging point. \par 
	In addition, diffraction calculations were conducted in the short-wavelength region. The calculations for \{250 MeV, 10 $\mu$m\} and \{10 MeV, 10 $\mu$m\} are shown in Figs. 3(g--i) and 3(j--l), respectively. The diffraction effect was less noticeable at a wavelength of 10 $\mu$m, and the focus sizes at the EO crystal were smaller. With an electron energy of 250 MeV, certain diffraction patterns remained. The diffraction patterns were invisible for an electron energy of 10 MeV because the effective electron radius was only $\sim$ 200 $\mu$m.  \par

	\subsubsection{Analytic model based on Fraunhofer assumption}
	
	Although a detailed calculation based on the Huygens--Fresnel principle can provide relatively accurate diffraction results for an arbitrary setup, the process becomes time-consuming when the number of frequency components exceeds a few hundred and the electron energy spectra are broad. Thus, we examined the feasibility of using a simplified analytical model based on the Fraunhofer assumption. The derivation process is described in \cite{xiang2007theoretical,CastellanoOTR}. To perform a quantitative calculation of the absolute field strength, we derived the TR field produced by a single electron at the imaging point in the SI unit as follows:
	\begin{equation}\label{FraunXiang}
		E_0^D(X_D, Y_D)=-\frac{e}{\epsilon_0}\frac{1}{\lambda M v}\frac{(X_D,Y_D)}{\rho_D}f(\theta_m,\gamma,\zeta)
	\end{equation}
	\noindent
	where $\rho_D=\sqrt{X_D^2+Y_D^2}$. $\theta_m = R_l/F$ is the acceptance angle of the OAP and $\zeta=k\rho_D/M$ is the normalized distance from the center of the TR field to the calculation point in the detector plane. Further, $M=F_2/F_1$ is the magnification of the imaging system. In this study, $M$ = 1. $f(\theta_m,\gamma,\zeta)=\int_{0}^{\theta_m}\theta^2/(\theta^2+\gamma^{-2})J_1(\zeta \theta)d\theta$ is a diffraction function defined by the optical system and the electron energy, where $J$ is a Bessel function of the first kind. This integration can be performed numerically. When $1/\gamma \ll \theta_m$, it is further simplified as follows \cite{xiang2007theoretical}:
	
	\begin{equation}\label{diffractionfunction}
		f(\theta_m,\gamma,\zeta)\approx\zeta^{-1}[\gamma^{-1}\zeta K_1(\gamma^{-1} \zeta)-J_0(\zeta \theta_m)]
	\end{equation}
	\noindent
	The electric field formed by an electron slice was then calculated as follows:
	\begin{equation}\label{Fraunslice}
		E^D(X_D,Y_D)=g_\perp \ast E_0^D
	\end{equation}
	
	The following assumptions were made to obtain  Eqs.~(\ref{FraunXiang},\ref{diffractionfunction}): (i) Fraunhofer approximation. The quadratic phase term in the source plane $\exp[ik(X_s^2+Y_s^2)/2F]$ was neglected, and the metal foil size was infinite. (ii) The distance between OAP1 and OAP2 is zero. An ideal thin lens was assumed, and intermediate diffraction was ignored. (An explanation for why the 2-OAP system can be geometrically treated as a thin-lens imaging system has been provided in Appendix B). (iii) The diameter of OAP1 acts as the only pupil in the system. (iv) The electrons possess high energy to satisfy the condition $1/\gamma \ll \theta_m$. It is expected that the field strength values from the detailed Huygens--Fresnel calculation should be smaller than those from the simplified model, wherein several diffraction processes are not considered. Therefore, the errors generated in this approach should be assessed. \par
	The calculation results for the same electron parameters as those in the previous section are shown in Fig. 4. Comparisons of Figs. 3(c), 3(i), and 3(l) with Figs. 4(a), 4(b), and 4(c), respectively, revealed that the focus shapes are almost identical over a large range of electron energies and wavelengths, though small errors exist. As our prediction, the peak values from the Huygens--Fresnel calculation were smaller than those of the analytical model, with errors of (7\%, 0.47\%, and 5.5 \%) in the cases of (\{250 MeV, 100 $\mu$m\}, \{250 MeV, 10 $\mu$m\}, and \{10 MeV, 10 $\mu$m\}), respectively. The error in the case of \{250 MeV, 100 $\mu$m\} primarily originated from the diffraction loss in the source plane and propagation. Whereas, the error for \{10 MeV, 10 $\mu$m\} resulted from the assumption $1/\gamma \ll \theta_m$. In the high-energy and short-wavelength regions, the errors were ignored. \par
	
	\subsection{Calculation of the OTR image}
	Before proceeding with the 3D calculations, we explain the calculation method for OTR images. OTR was used to determine the electron transverse bunch size. In the optical range, the wavelength was excessively short, which resulted in the diffraction difference between the Huygens--Fresnel principle and the Fraunhofer assumption being unnoticeable. Therefore, a simplified analytical model was used. The OTR image is an integrated TR intensity with a dimension ``$z$". Therefore, the issue of coherence expressed by the form factor must be considered.  \par
	For a certain frequency, the TR field in the EO crystal is the sum of all electrons in the bunch, that is, the integration of the TR produced by all slices, as illustrated in Fig. 2. The TR field of one frequency component in the EO crystal is expressed as,
	\begin{multline}\label{EoneF}
		E^D= N\iint dx_bdy_bg_{\perp}(x_b,y_b) E_0^D(X_D-x_b, Y_D-y_b) \\
		\times \int dz_b g_{\parallel}(z_b) \exp(i\omega z_b/v)
	\end{multline}
	
	\begin{figure}[ht]
		\centering
		\includegraphics[width=8.6 cm]{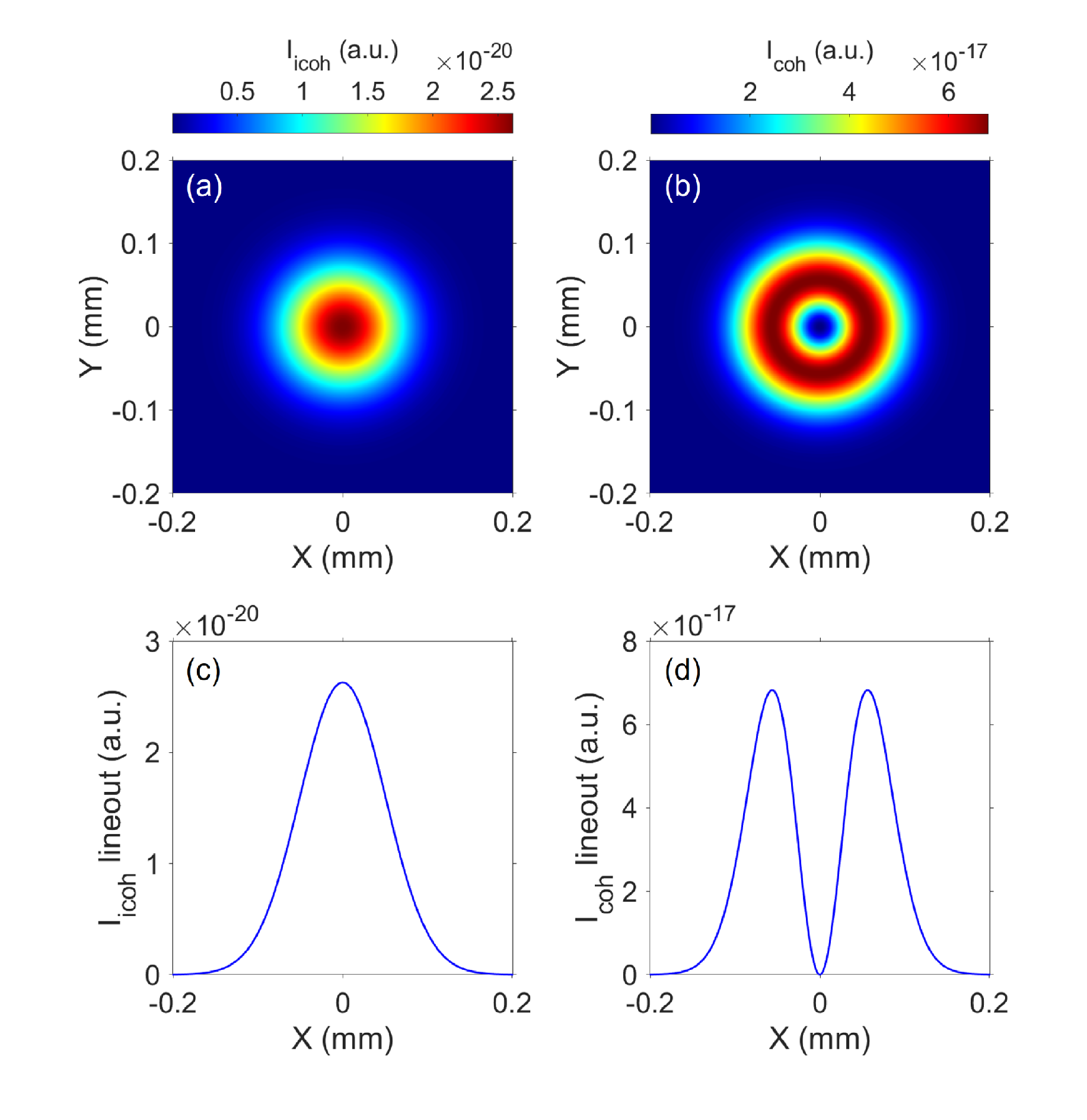}
		\caption{Calculation results of the OTR intensity distribution (arbitrary unit) at a wavelength of 800 nm. Initial transverse sizes of the Gaussian bunch are ($\sigma_x$, $\sigma_y$) = (50, 50) $\mu$m. The electron energy is 100 MeV and the charge is 32 pC. (a) and (b) show $I_{icoh}$ and $I_{coh}$ at an electron bunch duration of 1 fs. (c) and (d) show the line-outs along the $X$ axis with $Y$ = 0 in (a) and (b), respectively.} 
		\label{f5}
	\end{figure}
	\noindent
	
	Here, $E^D$ represents any \{x, y\} component of the TR field and $N$ is the number of electrons. The integration along the $z$ direction is the longitudinal form factor $F_z=\int dz_b g_{\parallel}(z_b) \exp(i\omega z_b/v)$, where $g_{\parallel}(z_b)$ is the longitudinal distribution of the bunch. The term $\exp(i\omega z_b/v)$ is given in Eq. (\ref{Erwxy22}) in Appendix A. The intensity of the OTR is $I_{OTR} = \sqrt{\epsilon_0/\mu_0}(|E_x^D|^2+|E_y^D|^2)/2$, which can be separated into an incoherent component $I_{icoh}$ and a coherent component $I_{coh}$ \cite{CTRschroeder} as follows: 
	
	\begin{equation}\label{OTRicoh}
		I_{icoh}=\sqrt{\epsilon_0/\mu_0}/2\times N\times [g_{\perp}\ast(|E_{0,x}^D|^2+|E_{0,y}^D|^2)]
	\end{equation}
	
	\begin{multline}\label{OTRcoh}
		I_{coh}=\sqrt{\epsilon_0/\mu_0}/2\times N(N-1)\times |F_z|^2 \\
		\times (|g_{\perp} \ast E_{0,x}^D|^2+|g_{\perp} \ast E_{0,y}^D|^2)
	\end{multline}
	
	\noindent
	where $\mu_0$ is the magnetic constant. The ratio of $I_{coh}/I_{icoh}$ is strongly dependent on the electron number and form factor. In the optical wavelength region, the ratio was sensitive when the electron bunch had a femtosecond or sub-femtosecond bunch duration. The calculation results for Gaussian bunches with a duration of 1 fs are shown in Fig. 5. As evident, $I_{icoh}$ exhibited a shape similar to the original electron bunch profile (Figs. 5(a) and 5(c)), whereas $I_{icoh}$ exhibited a ``donut" shape because of radial polarization and the symmetry of the electron transverse profile (Figs. 5(b) and 5(d)). In another calculation with a bunch duration of 2 fs, $I_{coh}$ decreased by a factor of $10^7$, whereas the change in $I_{icoh}$ was irrelevant with respect to the bunch duration. \par
	
	With the number of electrons set as $2\times10^8$ in this calculation, the peak value of $I_{coh}$ was $2.6\times10^3$ times higher than that of $I_{icoh}$. This indicates that, in the bunch, if there exists a femtosecond spike with 0.1\% of the total charge (tens to hundreds of fC), the overall OTR intensity profile exhibits a coherent pattern, as shown in Fig. 5(b). Such coherent OTR caused by femtosecond or sub-femtosecond spikes have been reported in both conventional accelerators \cite{loos2008observation} and LWFA \cite{cotrIlpa}. \par
	When the OTR was incoherent, deconvolution was conducted to retrieve the electron transverse profile by directly using the point spread function $|E_{0,x}^D|^2+|E_{0,y}^D|^2$, as explained in Eq. ~(\ref{OTRicoh}) and Ref. \cite{xiang2007theoretical}. However, when the OTR exhibits a coherent pattern, straightforward deconvolution is not possible. Thus, to estimate transverse electron sizes, an empirical linear relationship $X_{max}=A+B\sigma_x$ \cite{potylitsyn2020image} was discovered by evaluating the peak positions of the lineout, as shown in Fig. 5(d), where $A$ and $B$ are constants, $X_{max}$ is the peak position in the OTR signal, and $\sigma_x$ is the Gaussian bunch size. Therefore, to a certain extent, this method can be used to determine the electron bunch size. The reconstruction method for the fine structure of the original electron transverse profile from $I_{coh}$ will be investigated elsewhere. \par
	
	\begin{figure*}[ht]
		\centering
		\includegraphics[width=18 cm]{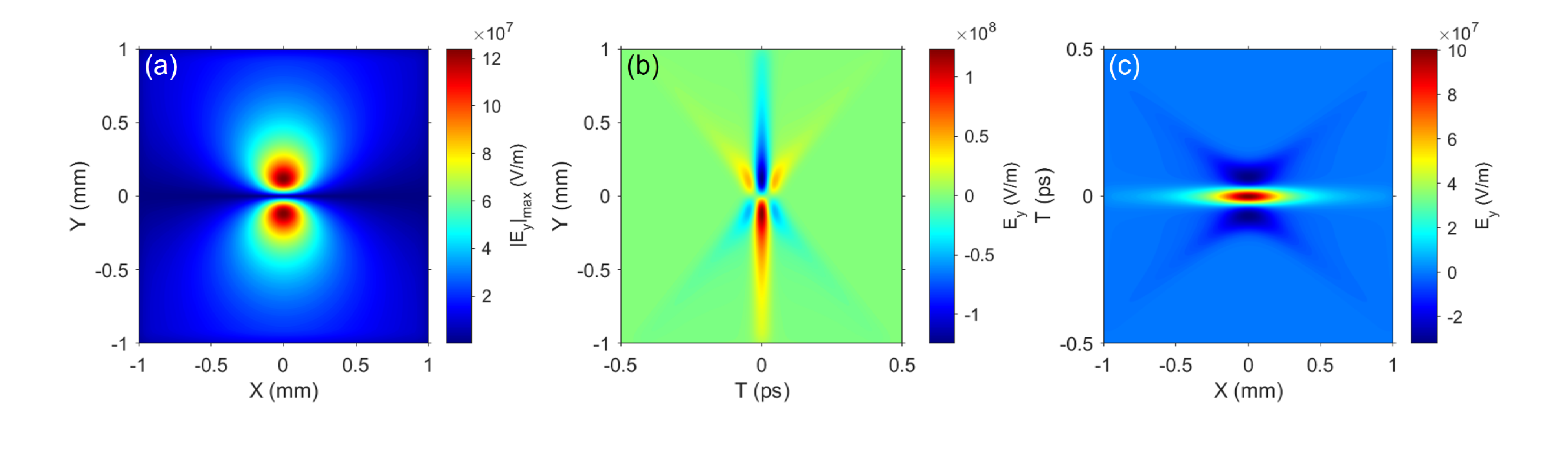}
		\caption{Calculation results of the 3D TR field $E_y(X,Y,T)$. (a) shows the ($X$, $Y$) distribution of the maximum value of $|E_y|$ in the third dimension $T$. (b) shows the ($T$,$Y$) distribution of $E_y$ at $X=0$. (c) shows the ($X$,$T$) distribution of $E_y$ at $Y=-200 \mu$m.} 
		\label{f6}
	\end{figure*}
	
	\subsection{3D TR field at the EO crystal}
	With a TR field of single frequency denoted by Eq. ~(\ref{EoneF}), the 3D TR field can be obtained using the Fourier transform as follows:
	\begin{equation}
		E(X_D,Y_D,t)=\frac{1}{2\pi}\int d\omega E^D(X_D,Y_D,\omega)\exp(-i\omega t)
	\end{equation}
	Calculations were performed with an electron energy of 100 MeV, a charge of 32 pC, and a bunch duration of 20 fs (rms). The size of the transverse bunch was (50, 50) $\mu$m. The calculations dealt with the time range of (-0.5--0.5) ps. The frequency range was zero to a few hundred THz. \par 
	To better explain the 3D (X,Y,T) profile, we plotted three 2D contours, as shown in Fig. 6. The spatial distribution of the maximum temporal TR value $|E_y|_{max}$ is shown in Fig. 6(a). The strong area of the TR field was concentrated within a few hundred micrometers. The size of the concentrated area was related to the electron energy and transverse bunch sizes. The $(T,Y)$ distribution of $E_y$ is shown in Fig. 6(b). The opposite values on both sides of $Y=0$ resulted from the radial polarization of TR. The EO signals were primarily determined by this distribution. The $(X,T)$ distribution of $E_y$ at $y=-200 \mu$m is shown in Fig. 6(c). The temporal profile of the TR field was not uniform along the $X$ axis. Previously, when experiments of EO spatial decoding were conducted, the transverse field distributions were not considered. However, when the magnification is not sufficiently large and the electron bunch has a small transverse size, the spatial strength profile of the field should be counted when uncompromising accuracy is required. \par
	
	\begin{figure}[ht]
		\centering
		\includegraphics[width=8.6 cm]{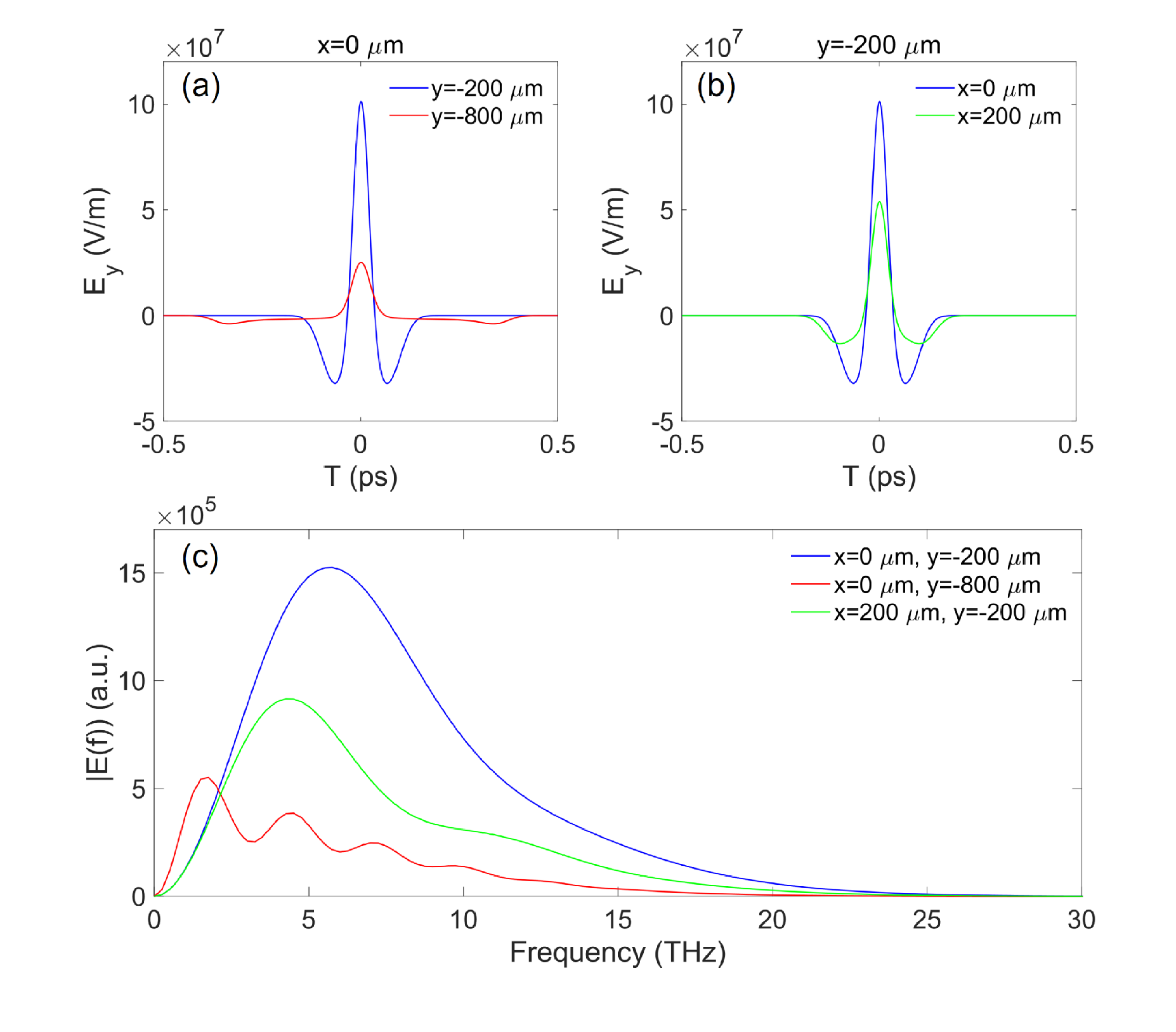}
		\caption{Lineouts of field profiles. (a) shows the temporal profiles at two vertical positions $Y=-200$ $\mu$m and $Y=-800$ $\mu$m in Fig. 6(b). (b) shows the temporal profiles at two horizontal positions $X=0$ $\mu$m and $X=200$ $\mu$m in Fig. 6(c). (c) shows the frequency domain profile of the TR field at three points (0, -200) $\mu$m (blue), (0, -800) $\mu$m (red), and (200, -200) $\mu$m (green).} 
		\label{f7}
	\end{figure}
	
	For a better elaboration of the spatial broadening of the TR field in the imaging plane, temporal and frequency domain lineouts were plotted, as shown in Fig. 7. The lineout plots for $Y=-200$ $\mu$m and $Y=-800$ $\mu$m illustrated in Fig. 6(b) are shown in Fig. 7(a). The lineouts at positions $x=0$ and $x=200$ $\mu$m are illustrated in Fig. 7(b). As evident, the field strength weakened and the duration was prolonged with an increase in the distance from the center of the field, both horizontally and vertically. The frequency distributions of the three points are shown in Fig. 7(c). The frequency distribution was narrower and shifted toward the long-wavelength end at greater distances from the center. The results shown in Fig. 7 indicate clear differences in the frequency compositions when measuring the field at different positions. Thus, the spatial distribution of the TR field should be considered for all forms of measurements, including temporal \cite{steffen2009electro} and frequency \cite{SchmidtTRspe} domain analyses.  \par
	
	\begin{figure}[ht]
		\centering
		\includegraphics[width=8.7 cm]{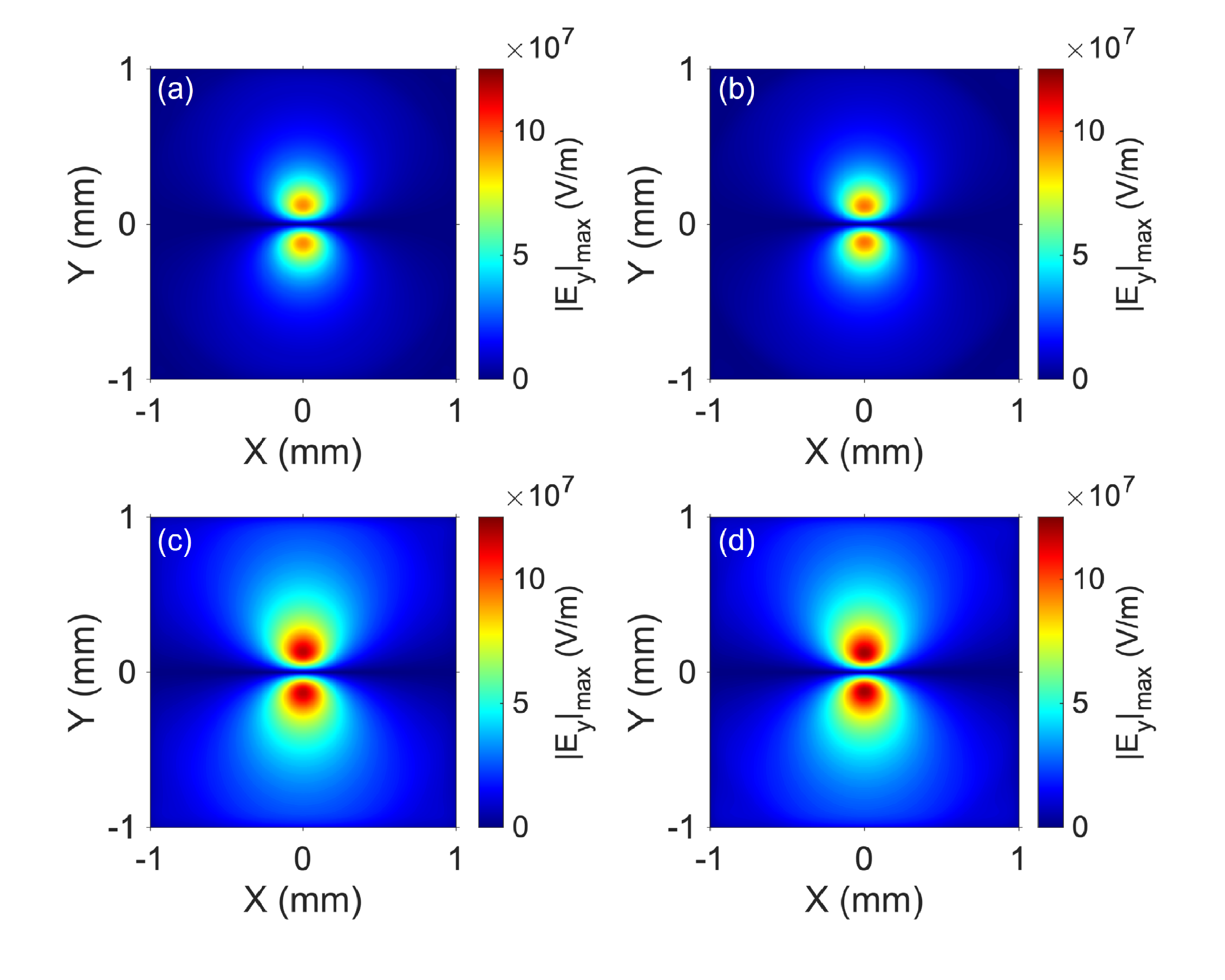}
		\caption{Comparison of the 3D calculations of the $|E_y|_{max}$ using the two approaches with different electron energies. All the figures share the same colormap limit. (a) and (c) show the calculations of Huygens--Fresnel diffractions. (b) and (d) show the calculations using the analytic model based on Fraunhofer assumption. (a) and (b) show the results with an electron energy of 20 MeV. (c) and (d) show the results with an electron energy of 200 MeV.} 
		\label{f8}
	\end{figure}
	
	\subsection{Comparison of 3D TR field calculations made using the two approaches}
	In Sections III.A and III.B, two approaches for calculating the image of the TR field: detailed diffraction based on the Huygens--Fresnel principle and an analytic model based on the Fraunhofer assumption for calculating, were elaborated. We observed slight differences in the monochromatic imaging calculations for various electron energies and wavelengths, as shown in Figs. 3 and 4. To confirm the possibility of adopting a less time-consuming approach in 3D calculations, the results of both approaches were compared.  \par
	The calculations from both approaches with different electron energies are plotted in Fig. 8. The electron bunch parameters are as follows: charge = 32 pC, transverse beam size = (50, 50) $\mu$m, and bunch duration = 20 fs (rms, gaussian). The TR field distributions obtained from both approaches were almost identical. With electron energies of both 20 and 200 MeV, the peak values of $|E_y|_{max}$ from the detailed calculations were only 3 \% lower than those of the simplified analytic model. The minute difference necessitates the use of an analytical model when performing parameter fitting. Based on the calculations above and in the previous subsections, we conclude that the analytical model is appropriate in most cases. In the lateral sections, without special announcements, an analytic model based on the Fraunhofer approximation will be applied in the calculation of the 3D TR field.  \par

	\section{Numerical study of EO spatial decoding}
	The fundamental inline EO sampling process has been studied numerically in detail in \cite{casalbuoni2008numerical}. Maximum phase retardation is achieved when the polarization of the probe laser, optical axis [-1,1,0] of the crystal, and polarization of the external field are in the same direction. 
	\begin{equation} \label{phaseretardationsimple}
		\Gamma=\frac{2\pi n_0^3d}{\lambda_0}r_{41}E_{TR}
	\end{equation}
	\noindent
	where $n_0$ is the refractive index of the central wavelength $\lambda_0$ of the probe laser, $r_{41}$ is the EO coefficient, and $d$ denotes the crystal thickness. Notably, Eq.~(\ref{phaseretardationsimple}) can only be used to approximately estimate the peak strength of the external field because the frequency-dependent response of the crystal is not included. Steffen et al. \cite{steffen2007electro} proposed a near cross-polarization setup for measuring the shape of a TR field with good linearity when $\Gamma$ was small. As shown in Fig. 1, a $\lambda/4$ plate was used to eliminate the residual birefringence caused by the EO crystal. The polarizer pair ``S" and ``P" are cross polarized to each other. The $\lambda/2$ plate was slightly rotated by an angle $\theta_2$, and the detected EO signal is calculated as, 
	\begin{equation}
		I_{det}(\theta_2,\Gamma)=\dfrac{I_p}{2}[1-\cos(\Gamma+4\theta_2)]+\delta_{ext}I_p+B_0
	\end{equation}
	\noindent
	where $B_0$ is the dark image of the camera, $\delta_{ext}$ is the extinction ratio of the polarizer pair, and $I_p$ is the original intensity profile of the probe. The background without an external TR field is expressed as follows:
	
	\begin{equation}
		B_1=\dfrac{I_p}{2}[1-\cos(4\theta_2)]+\delta_{ext}I_p+B_0
	\end{equation}
	\noindent
	We conducted data processing of $I_{sig}=(I_{det}-B_1)/(B_1-B_0)$ to eliminate the impact of nonuniformity of the original probe transverse profile as follows:
	\begin{equation} \label{EOsignalformNC}
		I_{sig}=\dfrac{\cos(4\theta_2)-\cos(\Gamma+4\theta_2)}{1-\cos(4\theta_2)+2\delta_{ext}}
	\end{equation}
	
	For EO spatial decoding, the detailed EO signal generation process has rarely been investigated outside the geometric temporal mapping relationship:
	\begin{equation} \label{TMrelationship}
		c\Delta \tau=\tan\theta_p \Delta \xi
	\end{equation}
	\noindent
	
	\begin{figure}[ht]
		\centering
		\includegraphics[width=8.6 cm]{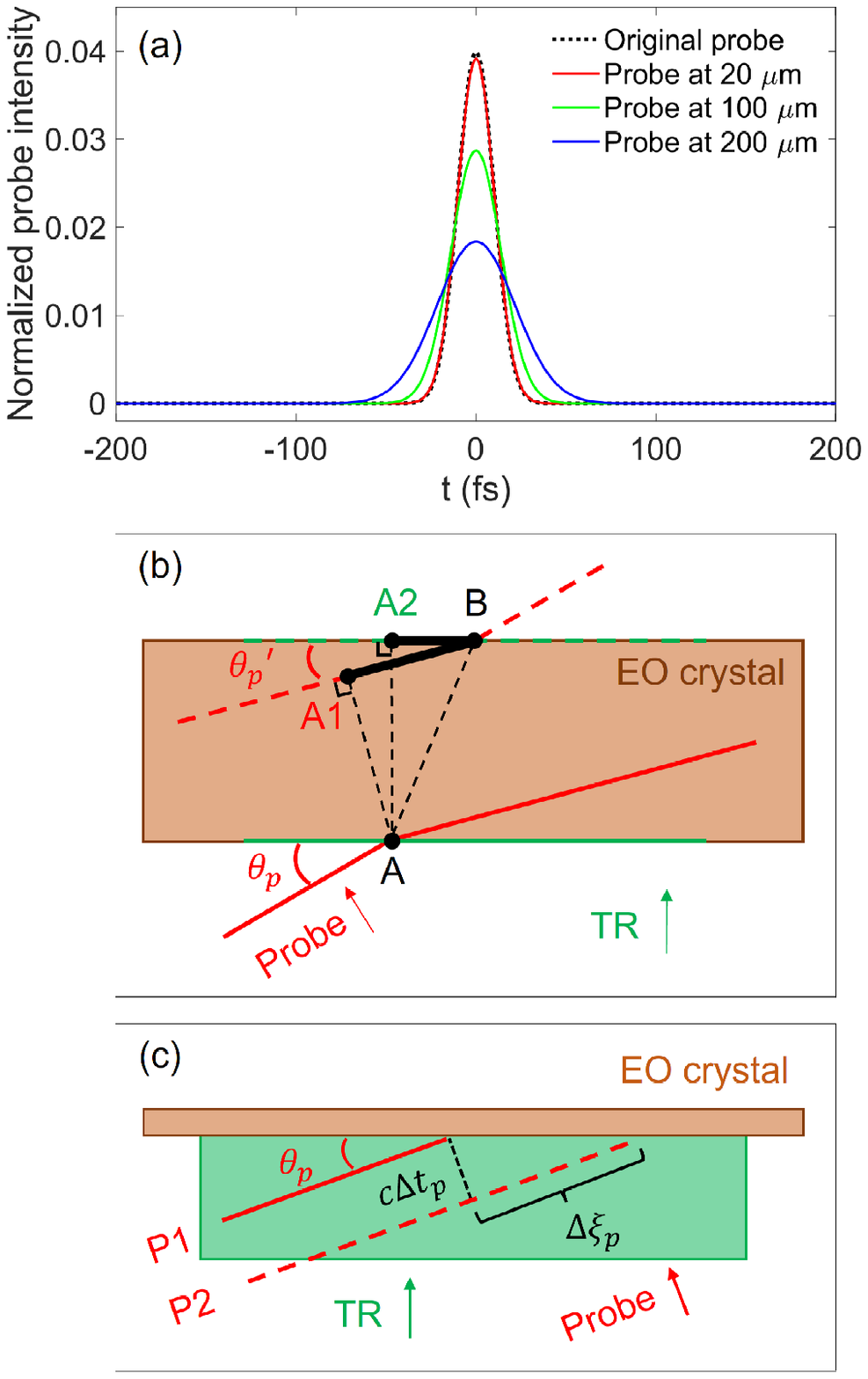}
		\caption{Geometric explanation of the EO spatial decoding process. (a) shows the normalized probe intensity profiles at propagation depths of 0 (dotted black), 20 $\mu$m (red), 100 $\mu$m (green), and 200 $\mu$m (blue). (b) illustrates the geometry of the walk-off between the TR field and the probe laser beam. The green and red solid lines show a timeline in the TR field and probe pulse when incident on a slice of the EO crystal, respectively. The dashed lines show those at the exit of the slice. The probe laser beam has incident and propagation angles of $\theta_p$ and $\theta_p'$, respectively. (c) shows how the existing probe duration should be considered in the calculation.} 
		\label{f9}
	\end{figure}
	
	where $\Delta \tau$ is the temporal gap, $\xi$ is the displacement observed by the camera, and $\theta_p$ denotes the incident angle of the probe laser beam on the surface of the EO crystal. When the dispersion of the TR field inside the crystal and the spatial distribution of the TR field are included, such a temporal mapping relationship lacks sufficient information to calculate the resulting EO signal shape. Thus, we conducted a detailed numerical study to apply EO spatial decoding to various situations. We arranged the elaboration from a simplified to a complicated case. The complexity of the calculation is dependent on the thickness of the crystal and spot size of the TR field in the image plane. Two main processes were involved in the calculation: \par
	(i) Temporal elongation of the probe laser beam: Here, we considered the use of gallium phosphide (GaP) crystals. The broadening effect can be calculated by considering the group delay dispersion (GDD) with the knowledge of the refractive index curve in the optical range $n(\lambda)=\sqrt{a_1+a_2\lambda^2/(\lambda^2-a_3)}$, where $a_1$, $a_2$, and $a_3$ are constants\cite{steffen2007electro}. The central wavelength of the probe laser beam was $\lambda_0$ = 800 nm. Assuming a Gaussian probe pulse with a duration of 10 fs (rms) and a flat phase profile in the frequency domain, the pulse shapes for various propagation depths are shown in Fig. 9(a). At a depth of 20 $\mu$m inside the GaP crystal, the shape exhibited minimal change. The probe pulse elongated upon further propagation inside the crystal. The curves in Fig. 9(a) were calculated with the integrated energy conserved. \par
	
	(ii) Walk-off between the probe laser beam and the TR field inside the EO crystal: The TR field was normally incident on a slice of the EO crystal. Owing to the relative angle and velocity differences between the TR field and the probe laser beam, a 2D geometry should be considered, as shown in Fig. 9(b). A timeline in the TR field propagated from the entrance to the exit of the crystal slice (solid green to dashed green). Within the same period, a timeline in the probe pulse propagated from the solid line to the dashed red line. As EO spatial decoding examines the transverse profile of the probe, the points ``A" and ``A1" shared the same timing on the camera. However, the cross point (coding point) shifted from ``A" to ``B”, which introduced walk-off in both the probe laser beam (denoted by the region $\overline{BA1}$) and the TR field (denoted by the region $\overline{BA2}$). $\overline{BA1}$ caused smearing of the EO signal in the transverse probe profile, and $\overline{BA2}$ is a factor that must be considered when the strength of the TR field changes rapidly in the transverse direction. \par

	\subsection{EO spatial decoding in a simple 1D case}
	
	\begin{figure*}[ht]
		\centering
		\includegraphics[width=17 cm]{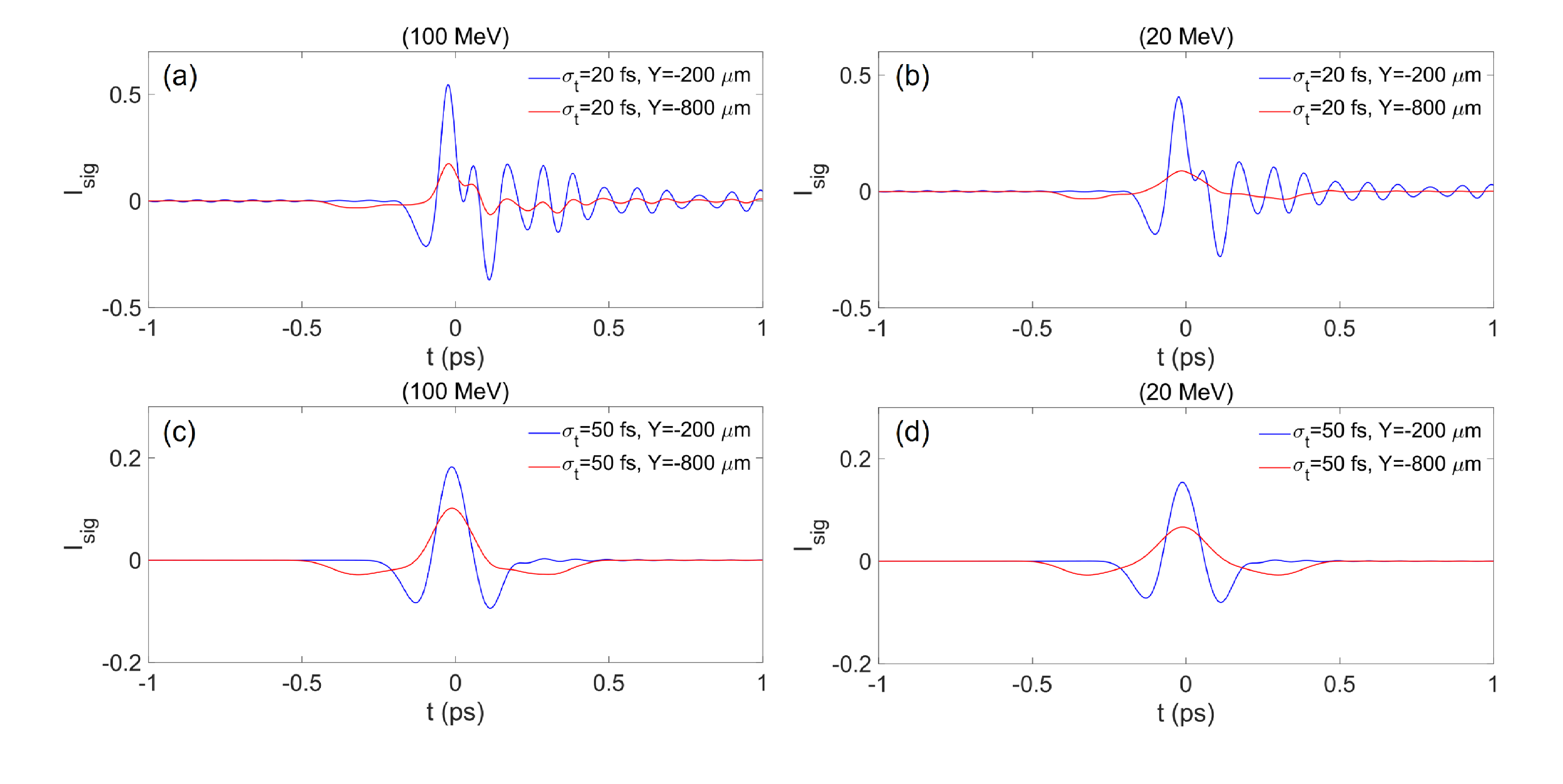}
		\caption{Calculated EO signals at points ($X$, $Y$) = (0, -200) $\mu$m (blue) and (0, -800) $\mu$m (red) using the near cross-polarization setup  with a 20-$\mu$m crystal. (a) and (b) show the results with electron bunch durations of 20 fs (rms). (c) and (d) show the results with electron bunch durations of 50 fs (rms). (a) and (c) correspond to an electron energy of 100 MeV. (b) and (d) show the results with an electron energy of 20 MeV.} 
		\label{f10}
	\end{figure*}

	As shown in Fig. 7, the TR fields in the temporal and frequency domains can be calculated at any spatial point on the imaging plane. The information regarding the frequency-dependent refractive index $N(\omega)=n(\omega)+i\kappa(\omega)$ and EO coefficient $r_{41}(\omega)$ can be found in \cite{casalbuoni2008numerical}. To further elaborate the process of EO signal generation, we separated the EO crystals into slices. Each slice $j$ had a thickness of $\Delta z$ and a propagation depth of $z_j$. \par
	The smearing process is described in \cite{Kaiprab} in the general case considering that the TR field has an incident angle of $\theta_s$. The smearing factor was calculated as follows:
	
	\begin{equation}\label{smearingSP}
		S=\dfrac{\sin\theta_p+\sin\theta_s}{\dfrac{v_p\sin\alpha\cos^2\theta_p'}{v_s-v_p\cos\alpha}+\sin\theta_p'\cos\theta_p'}
	\end{equation}
	
	\noindent
	where $\theta_p'$ and $\theta_s'$ are the propagation angles of the probe and the TR field inside the crystal, respectively, $\alpha=\theta_s'+\theta_p'$, $v_p$ denotes the group velocity of the probe, and $v_s = c/n(\omega)$ is the phase velocity of the TR field. Because $\theta_s=\theta_s'=0$, based on the relationship $\sin\theta_p=n_0\sin\theta_p'$, Eq.~(\ref{smearingSP}) is reduced to
	
	\begin{equation}\label{smearingP}
		S(\omega)=c(\frac{1}{v_\parallel}-\frac{1}{v_s})
	\end{equation}
	\noindent
	where $v_\parallel=v_p\cos\theta_p'$ denotes the probe speed component along the propagation direction of the TR field. For EO spatial decoding, the initial transverse profile of the probe laser beam was mapped as a time array $\tau$. In slice $j$, the time table changed owing to the walk-off $\overline{BA1}$ as follows:
	\begin{equation}
		t_j=\tau+S\times z_j/c
	\end{equation}
	Without considering the TR’s transverse distribution, we calculated only $E_{TR}^0(\omega)$ at a certain spatial point ($X$, $Y$) at the entrance of the crystal. At a propagation depth of $z_j$ inside the crystal, the TR field in the frequency domain was calculated as:
	\begin{equation}
		E_{TR,j}(\omega)=\frac{2}{1+N(\omega)}\exp[-\kappa(\omega)\omega z_j/c]E_{TR}^0(\omega)
	\end{equation} 
	as their own time references. Here, $2/(1+N(\omega))$ is the amplitude transmission coefficient based on the Fresnel's law and $\kappa(\omega)\omega/c$ is the absorption rate. The effective TR field was calculated by calculating the EO coefficient $r_{41}(\omega)$ as follows:
	\begin{equation}
		E_{TR,j}^{eff}(\omega)=r_{41}(\omega)E_{TR}(\omega)
	\end{equation}
	
	The EO effect was calculated in the time domain. The effective TR field $E_{TR,j}^{eff}(\tau)$ was obtained using Fourier transformation $E_{TR,j}^{eff}(\tau)=\frac{1}{2\pi}\int\exp(-i\omega t_j)E_{TR,j}^{eff}(\omega)d\omega$.
	\begin{widetext}
		\begin{equation}
			E_{TR,j}^{eff}(\tau)=\frac{1}{2\pi}\int \exp(-i\omega \tau)\exp[- i\omega z_j(\frac{1}{v_\parallel}-\frac{1}{v_s})]\frac{2}{1+N(\omega)}\exp[-\kappa(\omega)\omega z_j/c]E_{TR}^0(\omega)r_{41}(\omega)d\omega
		\end{equation}
	\end{widetext}
	The phase retardation in slice $j$ was then calculated as,
	\begin{equation}
		\Delta\Gamma_j(\tau)=\frac{2\pi n_0^3}{\lambda_0}\Delta z E_{TR,j}^{eff}(\tau)
	\end{equation}
	\noindent
	By setting the slice number to infinity, the overall phase retardation is the sum of the contributions from all slices $\Gamma(\tau)=\sum_{j}\Delta\Gamma_j(\tau)$ and can be transformed into an integral as follows:
	\begin{widetext}
		\begin{equation} \label{phaseretard1sliceprobe}
			\Gamma(\tau)= \frac{n_0^3d}{\lambda_0} \int d\omega \frac{2}{1+N(\omega)}E_{TR}^0(\omega)r_{41}(\omega) \times \frac{1}{d}\int_{0}^{d}\exp[i\omega(\frac{N(\omega)}{c}-\frac{1}{v_\parallel})z]dz
		\end{equation}
	\end{widetext} 
	
	\begin{figure}[ht]
		\centering
		\includegraphics[width=8.6 cm]{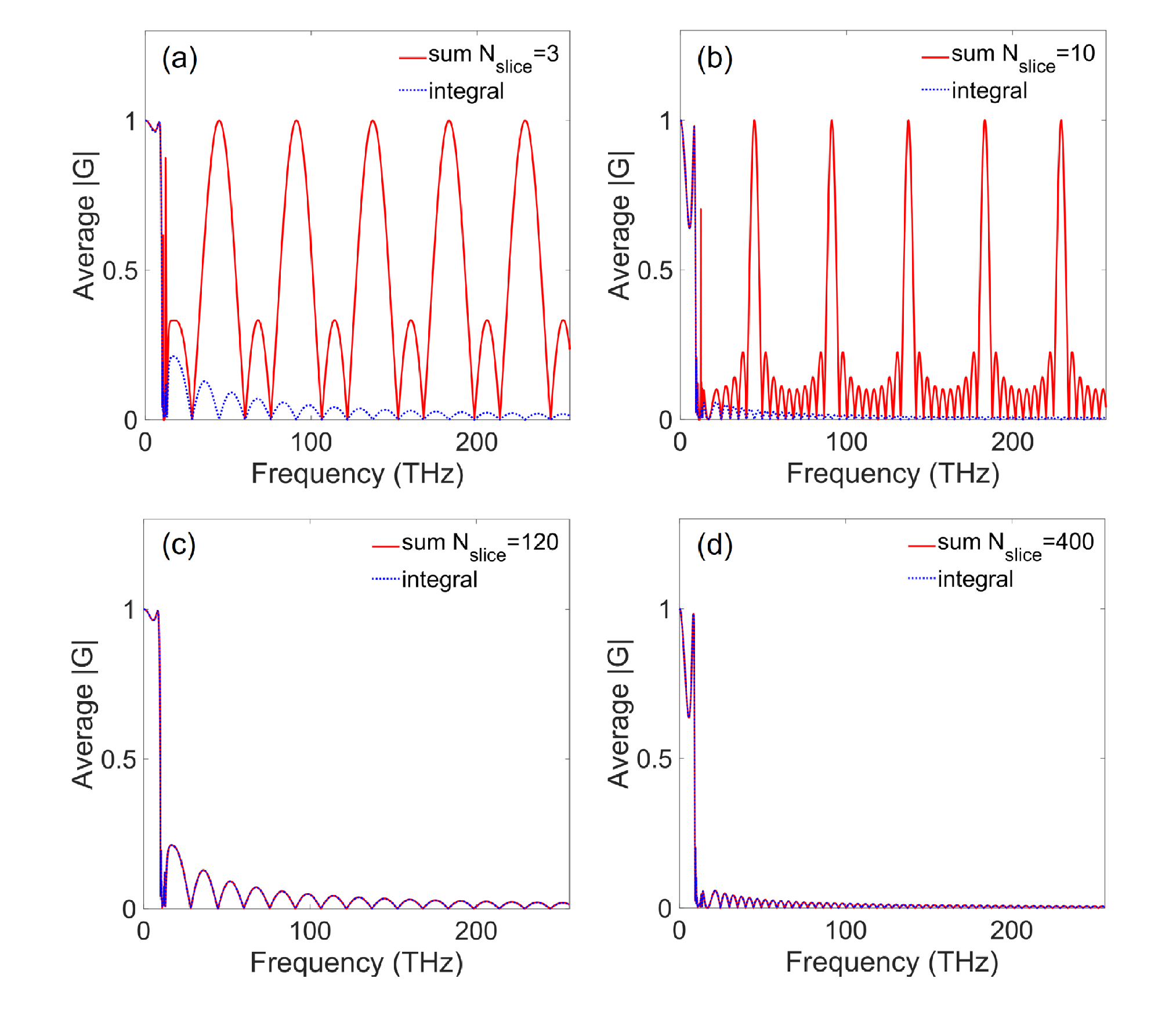}
		\caption{Averaged geometric response function with different number of slices. The results of $|G|_{integral}$ and $|G|_{sum}$ are denoted by blue dots and red curves, respectively. (a) and (b) correspond to cases with a small number of slices. (c) and (d) show the results from a large number of slices. (a) and (c) show the results from a 30-$\mu$m crystal. (b) and (d) show the results from a 100-$\mu$m crystal.} 
		\label{f11}
	\end{figure}
	
	\begin{figure}[ht]
		\centering
		\includegraphics[width=8.6 cm]{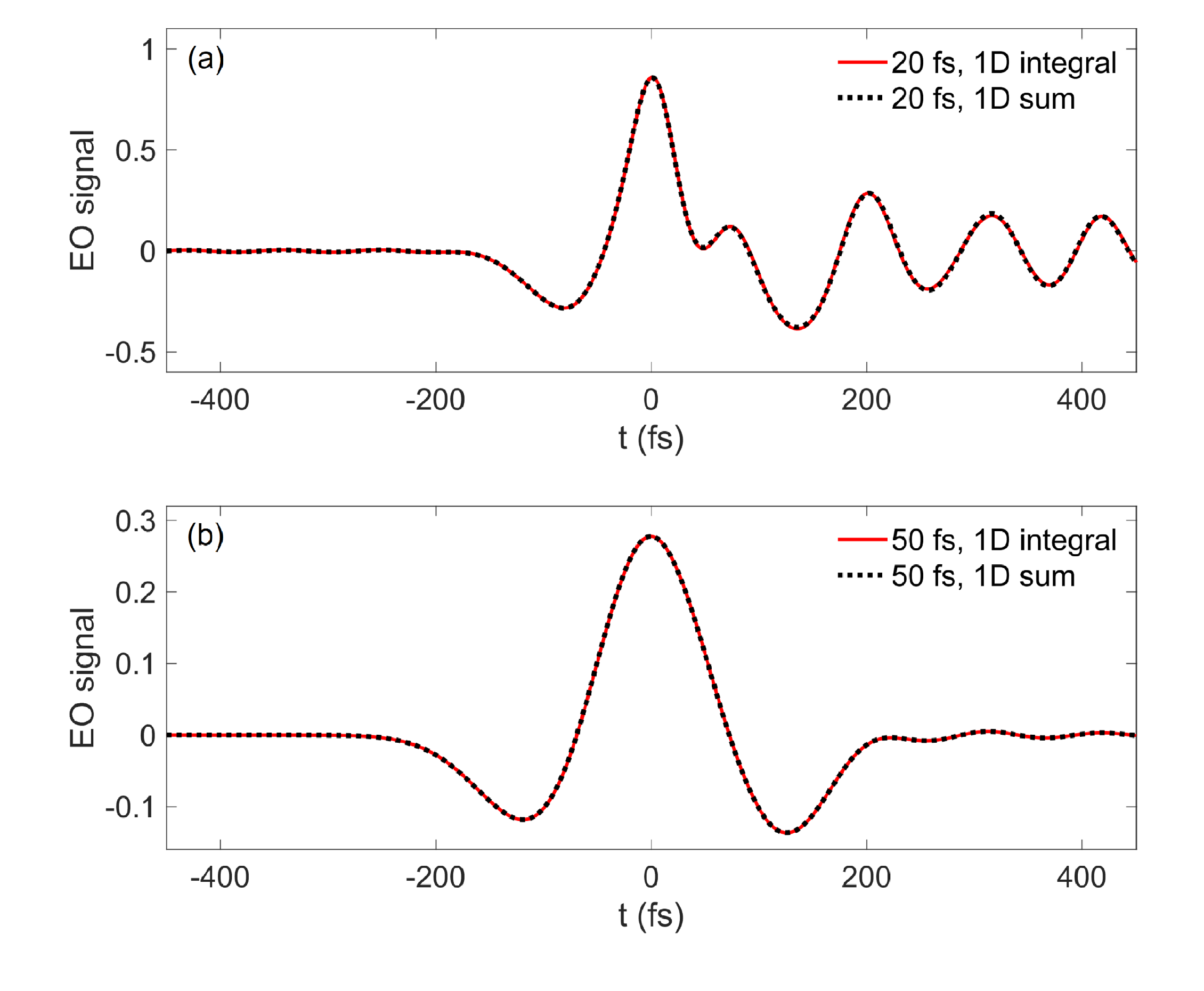}
		\caption{Comparison of the calculation results between the ``1D integral" and ``1D sum" models. The red curves and black dots denote the calculation results from the ``1D integral" and ``1D sum" models, respectively. (a) shows the results with a bunch duration of 20 fs. (b) shows the results with a bunch duration of 50 fs.} 
		\label{f12}
	\end{figure}

	Equation(\ref{phaseretard1sliceprobe}) expresses the phase retardation obtained using an infinitesimal slice of the probe laser beam. However, the probe laser beam had a longer pulse duration. We considered that the probe laser beam contained the maximum number of slices possible and calculated the overall phase retardation by summing all slices. This process is illustrated in Fig. 9(c). Two slices, denoted as ``P1" and ``P2", in the probe pulse had a time gap of $\Delta t_p$. In the probe laser beam with an incident angle of $\theta_p$, the same TR wavefront was encoded to the subsequent probe slice with a displacement of $\Delta \xi_p=c\Delta t_p/\tan\theta_p$ on the camera chip. When using the temporal mapping relationship in Eq.~(\ref{TMrelationship}), signal blurring occurred with a decoded time gap of $\Delta \tau_p=\tan\theta_p/c\times c\Delta t_p/\tan\theta_p=\Delta t_p$. \par
	Similarly, if the probe had temporal slices with an index $j$ and an intensity profile at time $I(t_p^j)$ in the time region between ($-T_p,T_p$), the total EO phase retardation is expressed as: $\Gamma_{total}(\tau)=\sum_{j}\Gamma(\tau+t_p^j)I(t_p^j)/\sum_{j}I(t_p^j)$. This was rewritten as an integral as follows:
	\begin{equation}\label{probeconv}
		\Gamma_{total}(\tau)=\int_{-T_p}^{T_p}dt_pf(t_p)\Gamma(\tau+t_p)
	\end{equation}
	\noindent
	where $f(t_p)$ denotes the normalized temporal intensity profile of the probe laser beam. Equation~(\ref{probeconv}) is in the form of a cross-correlation between the temporal profile of the probe laser beam and phase retardation of one probe slice, as denoted in Eq.~(\ref{phaseretard1sliceprobe}). Mathematically, when the kernel profile is symmetrical around zero, similar to that in the case of a Gaussian-like profile, Eq.(\ref{probeconv}) is equivalent to a convolution. \par
	Examples of the calculations are shown in Fig. 10 for electron energies of \{100, 20\} MeV and bunch durations of \{20, 50\} fs (rms). Frequency-domain TR fields at two points ($X$, $Y$) = (0, -200) $\mu$m and (0, -800) $\mu$m were used as inputs for EO signal generation. The electron transverse bunch sizes were (50, 50) $\mu$m, and Eq.~(\ref{EOsignalformNC}) was used to calculate $I_{sig}$. The probe laser beam was assumed to exhibit a Gaussian shape with a duration of 10 fs (rms), and the incident angle of the probe laser beam was $\theta_p$ = 25$\degree$. The oscillations in the $I_{sig}$ curves resulted from frequency-dependent phase mismatching. \par
	
	The following observations were obtained from Fig. 10. (i) The visibility of oscillations is strongly dependent on the electron bunch duration. Fierce oscillations are observed for electron bunching duration of 20 fs. For electrons with a bunch duration of $>$ 50 fs, the oscillations are indistinctive and the shapes of $I_{sig}$ are almost identical to the temporal shape of the TR field. (ii) The signals are broadened and weaker when the signal was measured at $Y$ = -800 $\mu$m. With a lower electron energy, the signal intensity further decreases. (iii) The signal intensity ratio of $I_{-200}/I_{-800}$ varies for different electron bunch durations and electron energies. The features above indicate that, to monitor the electron bunch duration or peak current, the spatial point at which the EO signal is measured must be identified. \par
	
	\begin{figure*}[ht]
		\centering
		\includegraphics[width=17.5 cm]{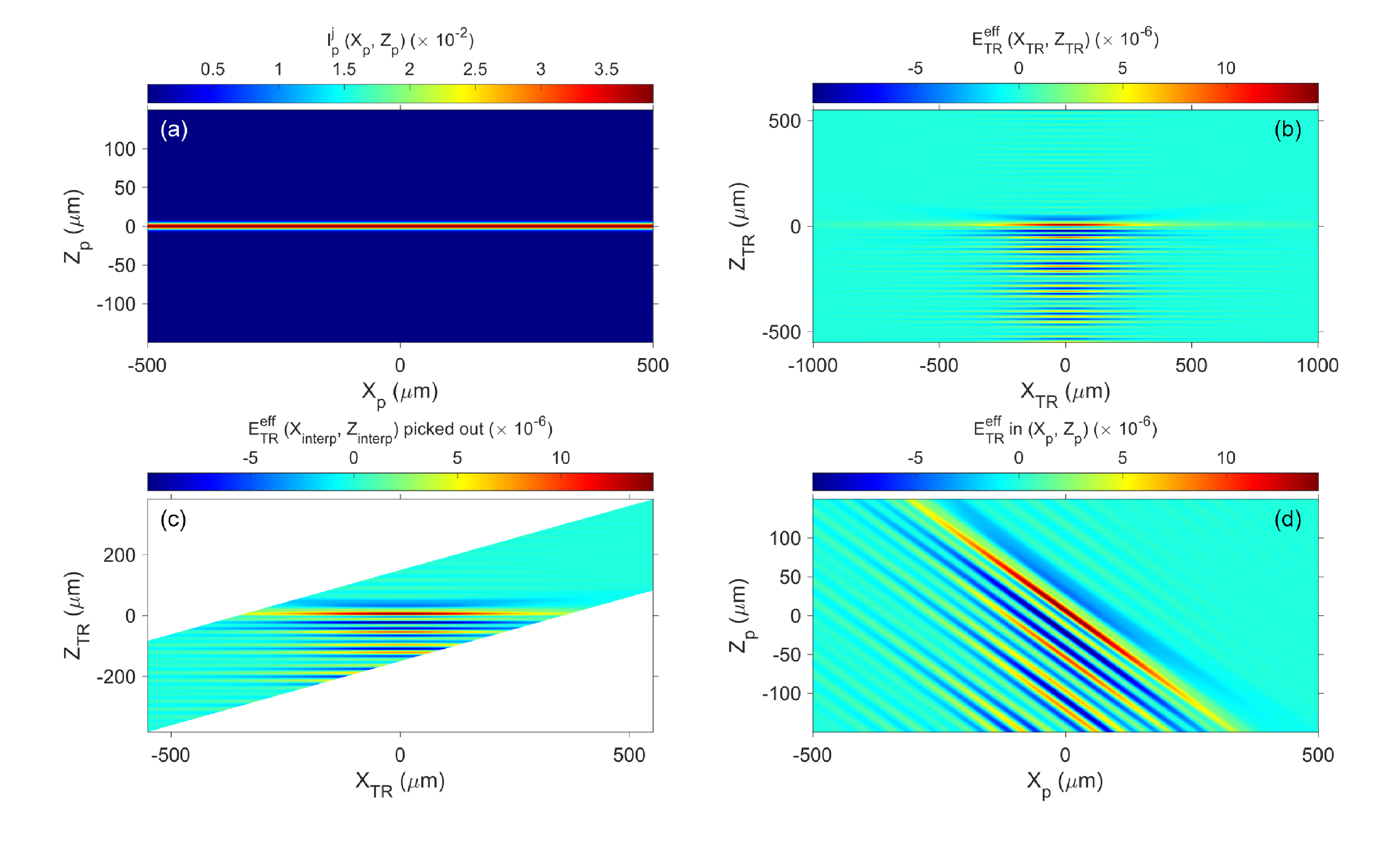}
		\caption{Schematic showing the calculation procedure for 2D EO spatial decoding. (a) shows the 2D probe intensity distribution at coordinates ($X_p$, $Z_p$). (b) shows the effective TR field inside the crystal at coordinates ($X_{TR}$, $Z_{TR}$). (c) shows the interpolated TR field at coordinates ($X_{TR}$, $Z_{TR}$). (d) shows the pattern of the interpolated effective TR field by coordinate transformation from ($X_{TR}$, $Z_{TR}$) to ($X_p$, $Z_p$).} 
		\label{f13}
	\end{figure*}

	\begin{figure}[ht]
		\centering
		\includegraphics[width=7 cm]{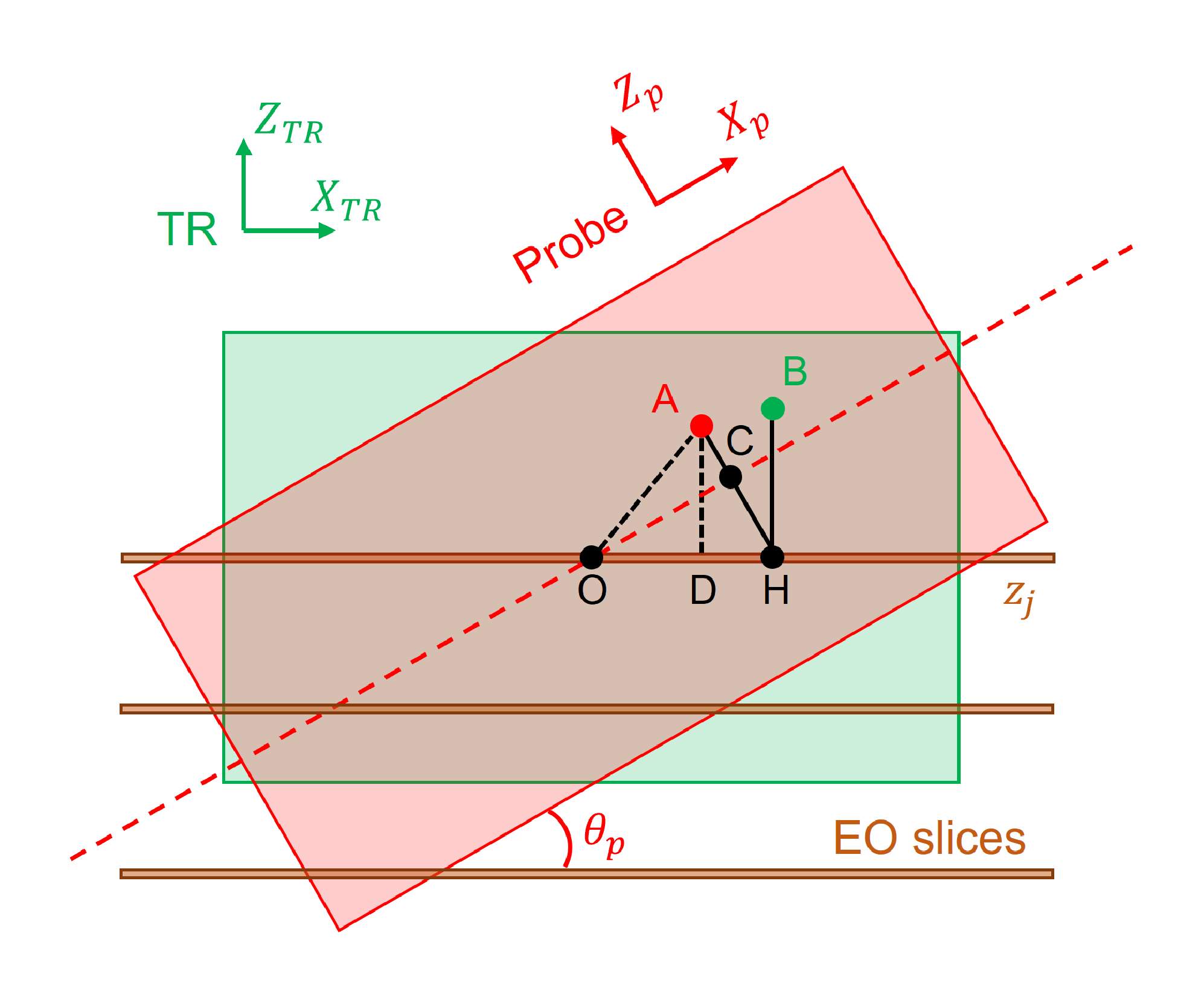}
		\caption{Schematic showing coordinate transformation when performing the 2D EO calculation. ``O" represents the origin of the coordinates. ``A" ($X_p,Z_p$) is a point in the probe laser. ``B" ($X_{interp},Z_{interp}$) is the point in the effective TR field meeting with point ``A" in the crystal slice $z_j$.  } 
		\label{f14}
	\end{figure}
	
	\begin{figure}[ht]
		\centering
		\includegraphics[width=8.5 cm]{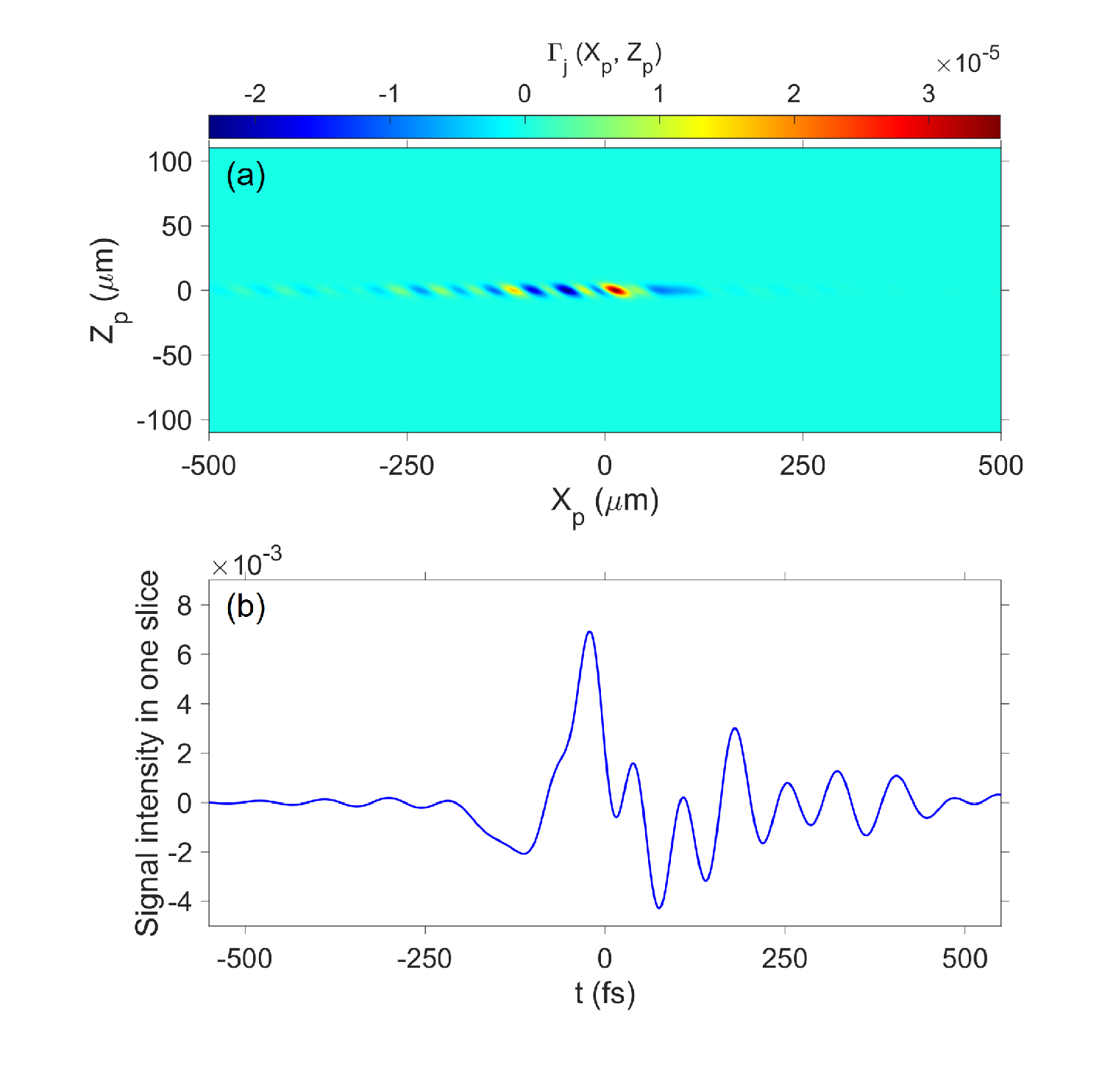}
		\caption{EO signal generation in one slice. (a) shows the 2D phase retardation distribution at coordinates ($X_p$,$Z_p$). (b) shows the accumulated EO signal in one slice in the near cross-polarization setup.} 
		\label{f15}
	\end{figure}

	\begin{figure}[ht]
		\centering
		\includegraphics[width=8.2 cm]{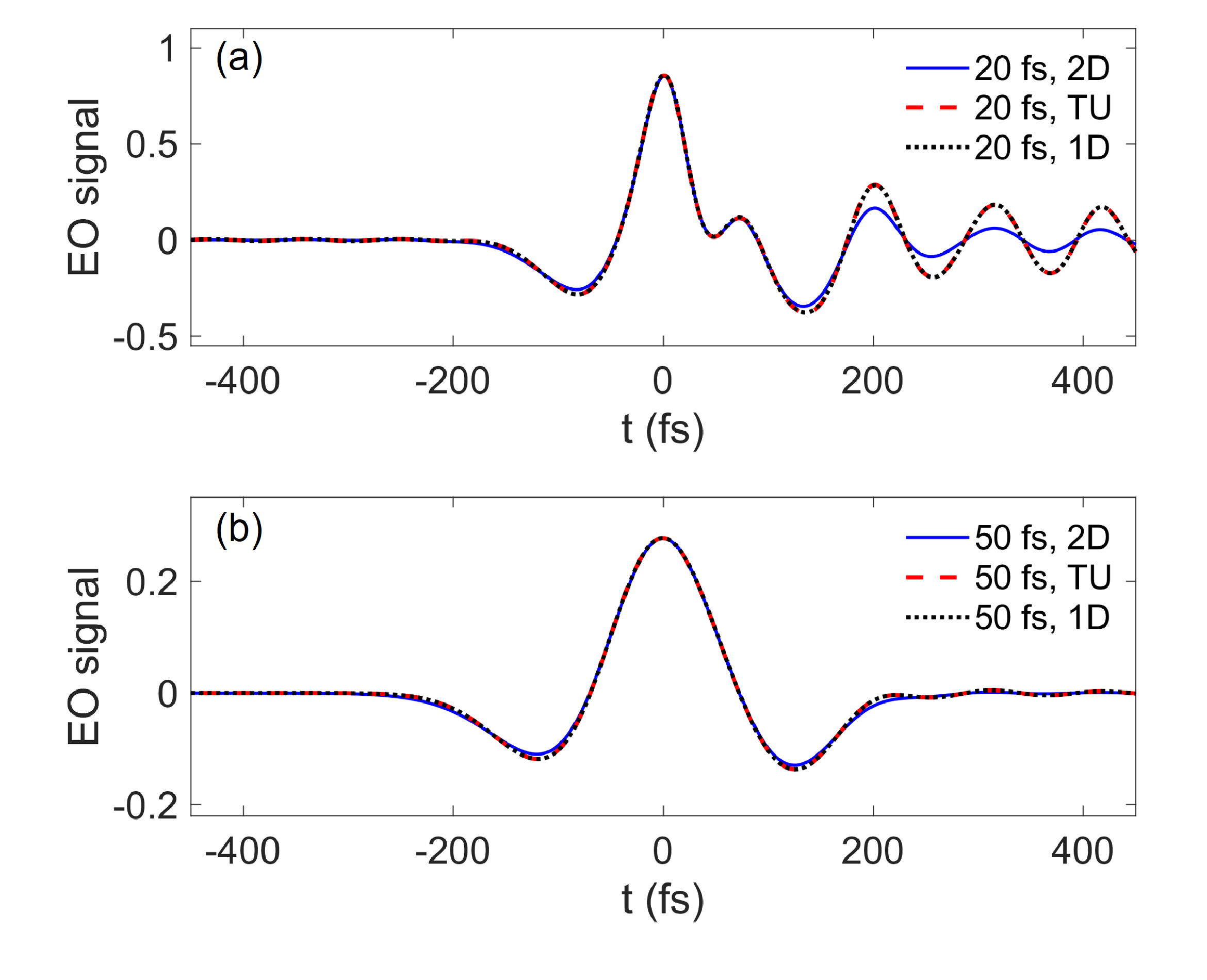}
		\caption{Overall EO signals generated from a GaP crystal with a thickness of 30 $\mu$m determined using different calculation models. The field information is obtained at $Y$ = -200 $\mu$m.  (a) shows the calculation results with an electron bunch duration of 20 fs (rms). (b) shows the calculation results with an electron bunch duration of 50 fs (rms). The results from the ``2D", ``TU" and ``1D" models are illustrated by blue solid curves, red dashed, and black dotted curves, respectively.}
		\label{f16}
	\end{figure}

	\subsubsection{Reminder on the adequate number of slices in EO sampling calculations}
	
	To manage the temporal elongation of the probe laser beam in a thick crystal, the EO crystal was separated into slices, as described above. A similar procedure was followed in \cite{steffen2007electro,casalbuoni2008numerical}. Here, the number of slices must be focused upon. We proceeded with this explanation by introducing a parameter that measures the phase mismatching process: the averaged geometric response function $|G|$. The $|G|$ value of the two approaches is calculated as
	
	\begin{eqnarray}\label{geointegral}
		|G|_{integral}=\frac{1}{d}|\int_{0}^{d}\exp(i\Delta k z) dz|
	\end{eqnarray} 
	
	\begin{eqnarray}\label{geosum}
		|G|_{sum}=\frac{1}{d}|\sum_{j}\exp(i\Delta k z_j)\times (d/N_{slice})|
	\end{eqnarray} 
	
	\noindent
	Equation~(\ref{geointegral}) is the final integral of Eq. ~(\ref{phaseretard1sliceprobe}) for a thin crystal and $\Delta k(\omega)=\omega(\frac{N(\omega)}{c}-\frac{1}{v_\parallel})$. Notably, Eq.~(\ref{geointegral}) is a modification of the inline version reported in Ref. \cite{casalbuoni2008numerical}, which was obtained by changing the group velocity of the probe $v_p$ to $v_\parallel$, considering the relative angle between the probe and the TR field. However, for thick crystals, it is inadequate to consider the entire crystal as a single slice. The average geometric response function is the sum of phase mismatches in each slice, as described in Eq.~(\ref{geosum}), where $N_{slice}$ denotes the number of slices used in the calculations and $d$ denotes overall crystal thickness. \par
	Figure 11 shows the results of the calculations for GaP crystals with thicknesses of 30 $\mu$m (Figs. 11(a) and 11(c)) and 100 $\mu$m (Figs. 11(b) and 11(d)). Cases with small $N_{slice}$ had a slice thickness of 10 $\mu$m, whereas cases with large $N_{slice}$ had slice thickness of 0.25 $\mu$m. When the number of slices was insufficient, $|G|_{sum}$ differed significantly from $|G|_{integral}$, which resulted in fake responses in the high-frequency region. This comparison indicates that, to achieve an appropriate shape of the EO signal, the number of slices must be increased.  \par
	Using the information described above, we carefully examined the validity of ``1D integral" calculations in a slightly thicker crystal with a thickness of 30 $\mu$m. In this crystal, the probe laser beam was slightly elongated in the last 10 $\mu$m. Figure 12 shows the results of the comparison between the ``1D integral" and ``1D sum" calculations. A total of 120 $N_{slice}$ were utilized in the ``1D sum" calculation and the probe elongation in each slice was calculated. In the ``1D integral" calculation, the entire crystal was considered as a single slice. An unchanged probe shape was obtained by calculating the average probe profile of 120 slices. At electron bunch durations of 20 and 50 fs, no differences were observed between the two models. This conclusion is essential in the analysis of the experimental EO signal because the calculation with many slices is two orders of magnitude slower than the ``1D integral" calculation. However, as this study utilized the numerical methodology, the notation ``1D model" implies the ``1D sum" model in lateral context for accuracy. \par

	\subsection{EO spatial decoding in a general 2D case}
	
	In this study, the transverse TR strength distribution was ignored, as described. However, when the temporal scale of interest is longer, a larger transverse region must be calculated, where the transverse TR field distribution may have an impact. Here, the TR field distribution plotted in Fig. 6(c) was used as the input. \par
	The EO crystal was divided into thin slices. For example, a GaP crystal with a thickness of 30 $\mu$m was separated into 120 slices. Each slice had a thickness of $\Delta z$ = 0.25 $\mu$m. The propagation depth at each slice was $z_j=j\Delta z$. The effective 2D TR fields $E_{TR}^{eff, j}(X_{TR},Z_{TR})$ in the time domain and the 2D probe intensities $I_p^j(X_p,Z_p)$ were calculated for each slice. Here, $Z$ and $Z_p$ are $Z_{TR}=-ct_{TR}$ and $Z=-ct_p$, respectively. further, $(X_{TR}, Z_{TR})$, and $(X_p, Z_p)$ are the coordinates that co-propagated with the TR and probe laser beams. Notably, the calculation of $E_{TR}^{eff, j}(X_{TR},Z_{TR})$ already included the EO effect and longitudinal phase mismatch. \par

	The calculation procedure for one slice of the EO crystal is illustrated in Fig. 13. The propagation depth was $z_j$ = 25 $\mu$m. The 2D intensity distribution of the probe laser beam with coordinates ($X_p$, $Z_p$) is shown in Fig. 13(a). Because the transverse intensity distribution of the probe was excluded using Eq.~(\ref{EOsignalformNC}), the transverse profile of the probe along $X_p$ was set to be uniform. Further, the probe intensity was normalized in the $Z_p$ ($t_p$) direction and calculated, including the GDD. The effective 2D TR field in the same slice is shown in Fig. 13(b). \par
	
	A mesh grid for interpolation was calculated for each slice. We wish to clarify that coordinate transformation is not simply a rotation of ($X_p$, $Z_p$). The relative propagation between the TR pulse and the probe laser beam should be considered for the EO effect occurring in a certain slice $z_j$. The geometry is shown in Fig. 14. By including the transverse shift and longitudinal phase mismatch, the two coordinates ($X_p$, $Z_p$) and ($X_{TR}$, $Z_{TR}$) were set to have the same origin, O (0, 0). Point ``H" denotes the point where the EO effect occurred at an earlier timing, contributed by point ``A" in the probe laser beam and point ``B" in the effective TR field. Thus, we obtained the relationship $\overline{HA} = \overline{HB}$. When $\overline{OC} = X_p$ and $\overline{AC} = Z_p$, we showed that $\overline{OH} = X_p/\cos\theta_p$ and $\overline{HA} = \overline{HB} = X_p \cdot \tan \theta_p + Z_p$. Furthermore, by subtracting the relative transverse shift $d\cdot\tan\theta_p'$, the interpolation coordinates were expressed using Eq.~{\ref{interpmesh}}. \par
	
	\begin{equation}\label{interpmesh}
		\left\{ 
		\begin{alignedat}{2}   
			X_{interp}&=X_p/\cos\theta_p-d\cdot\tan\theta_p' \\ 
			Z_{interp}&=X_p\tan\theta_p+Z_p
		\end{alignedat} 
		\right.
	\end{equation}
	
	\noindent
	where $\theta_p'$ denotes the propagation angle of the probe beam inside the crystal. Subsequently, a 2D interpolation of $E_{TR}^{eff, j}(X_{interp},Z_{interp})$ was conducted in the region determined using Eq.~(\ref{interpmesh}). The interpolated TR field distribution is illustrated in Fig. 13(c). Next, we transformed $E_{TR}^{eff, j}(X_{interp}, Z_{interp})$ into the probe laser coordinates ($X_p$, $Z_p$), as shown in Fig. 13(d). \par  
	
	Subsequently, we calculated the phase retardation in each pixel as $\Gamma_j(X_p, Z_p)=2\pi n_0^3 \Delta d/\lambda_0 \times I_p^j(X_p,Z_p)E_{TR}^{eff, j}(X_p, Z_p)$, which is the product of Fig. 13(a) and 13(d). A 2D distribution of $\Gamma$ is shown in Fig. 15(a). The phase retardation profile $\Gamma_j(X_p)$ along $X_p$ was then calculated by accumulating the results shown in Fig. 15(a) in the $Z_p$ direction. The EO signal contributed by this slice was calculated using Eq.~(\ref{EOsignalformNC}). Earlier timing corresponded to larger $Z_{TR}$ and $X_{p}$ in Fig. 14. The timetable was arranged as $t=-X_p\tan\theta_p/c$. The EO signal from one slice is shown in Fig. 15(b). The peak position of the signal was at a timing earlier than zero because the group velocity of the probe laser beam was smaller than the phase velocity of the TR field inside the EO crystal. In this article, without special clarification, the peak of the EO signal is shifted to ``zero" timing.\par
	
	The information described above indicates the way EO signal generation is observed within one slice in a 2D geometry. The overall EO signal was calculated in three steps: (i) calculation of the accumulated phase retardation $\Gamma_j(X_p)$ in each slice, (ii) calculation of the overall phase retardation as $\Gamma(X_p)=\sum_{j}\Gamma_j(X_p)$, and (iii) calculation of the EO signal using Eq.~(\ref{EOsignalformNC}). \par
	
	Figure 16 shows the 2D EO spatial decoding results at the vertical position $Y$ = -200 $\mu$m. We performed the calculations with three different models: (i) ``2D" (blue curve) implies the 2D EO calculation by considering the transverse TR field strength distribution; (ii) ``TU" (red dot) implies the 2D EO calculation assuming that the TR field is transversely uniform; and (iii) ``1D" (black dashed line) denotes the analytical model described in Sec. IV. A. $\Gamma_j$ was calculated for each slice using Eqs. ~(\ref{phaseretard1sliceprobe}) and ~(\ref{probeconv}). \par
	
	The calculations revealed interesting results. (i) The results from ``TU" and ``1D" were identical, indicating that our approach for the 2D EO calculation is adequate. (ii) For a slightly longer bunch duration of 50 fs, the difference between the ``2D" and ``TU" was unnoticeable. (iii) For a very short bunch duration of 20 fs, determining a difference around the main peak of the signal was challenging. However, the amplitudes of the oscillations at later times were slightly smaller. Whereas, for a very long electron bunch, although no calculations were performed, we expected visible differences between the ``2D" and ``1D" models. In the case of LWFA, as the electron bunches have duration of a few fs to tens of fs, with such small errors, the application of the ``1D" model should provide sufficient results when analyzing the EO signal around the main peak.  \par

	\subsection{EO spatial decoding signal of temporally chirped electron bunches}
	
	Using the calculation method established above, we investigated electron bunches with multiple energy components. Typically, electron bunches with energy spreads exhibit chirps in the longitudinal phase space (LPS). We addressed this issue by separating the electrons into energy slices corresponding to different timings. For simplicity, a linear energy chirp with an energy range of 20–300 MeV was assumed. For electron slice j, the timing was $\delta t_j=chirp\cdot(E_j-160)$ and the charge weight was $\eta_j$. The longitudinal form factor was calculated using the Fourier transform $F_z^j(\omega)=\int \exp(i\omega t) \times \exp[-(t-\delta t_j)^2/2\sigma_{t0}^2]dt$. The result of this integral was obtained as
	\begin{equation} \label{multibunch}
		F_z^j(\omega)= \exp(-\frac{\omega^2\sigma_{z0}^2}{2v^2}+\frac{i\omega\delta z_j}{v})
	\end{equation}
	
	\noindent
	for relativistic electrons $v\approx c$. Each slice was assumed to have the same slicing bunch duration of $\sigma_{t0}$  = 10 fs, corresponding to a length of $\sigma_{z0}=c\sigma_{t0}$. With the relative center timing of each slice set to $\delta t_j$, we defined $\delta z_j = c\delta t_j$. We assessed the signals from ``2D" EO calculations at a vertical distance of $Y_D=-200$ $\mu$m. The electron energy spectrum of a two-temperature (``2T") distribution $dN/dE \propto 1/T_1 \exp(-E/T_1)+1/T_2\exp(-E/T_2)$ was evaluated, with $T_1$ = 20 MeV and $T_2$ = 300 MeV. Examples of temporally chirped electron bunches are shown in Fig. 17. Figure 17(a) shows the energy distributions (blue) of the ``2T" model and a temporal chirp (red) of 0.2 fs/MeV. The 2D plot of the LPS and the current of a two-temperature electron bunch with a chirp of 0.2 fs /MeV is illustrated in Fig. 17(b). Using the field expressions presented in Eq.~(\ref{FraunXiang}), the frequency-domain field component is the sum of contributions from each electron energy component and the corresponding $F_z^j(\omega)$, and is calculated as follows:
	
	\begin{equation}\label{gammasumfield}
		E^D(X_D,Y_D,\omega)=\sum_{j}\eta_j E_0^D(X_D,Y_D,\gamma_j,\omega)F_z^j(\omega)
	\end{equation}
	
	\begin{figure}[ht]
		\centering
		\includegraphics[width=8.6 cm]{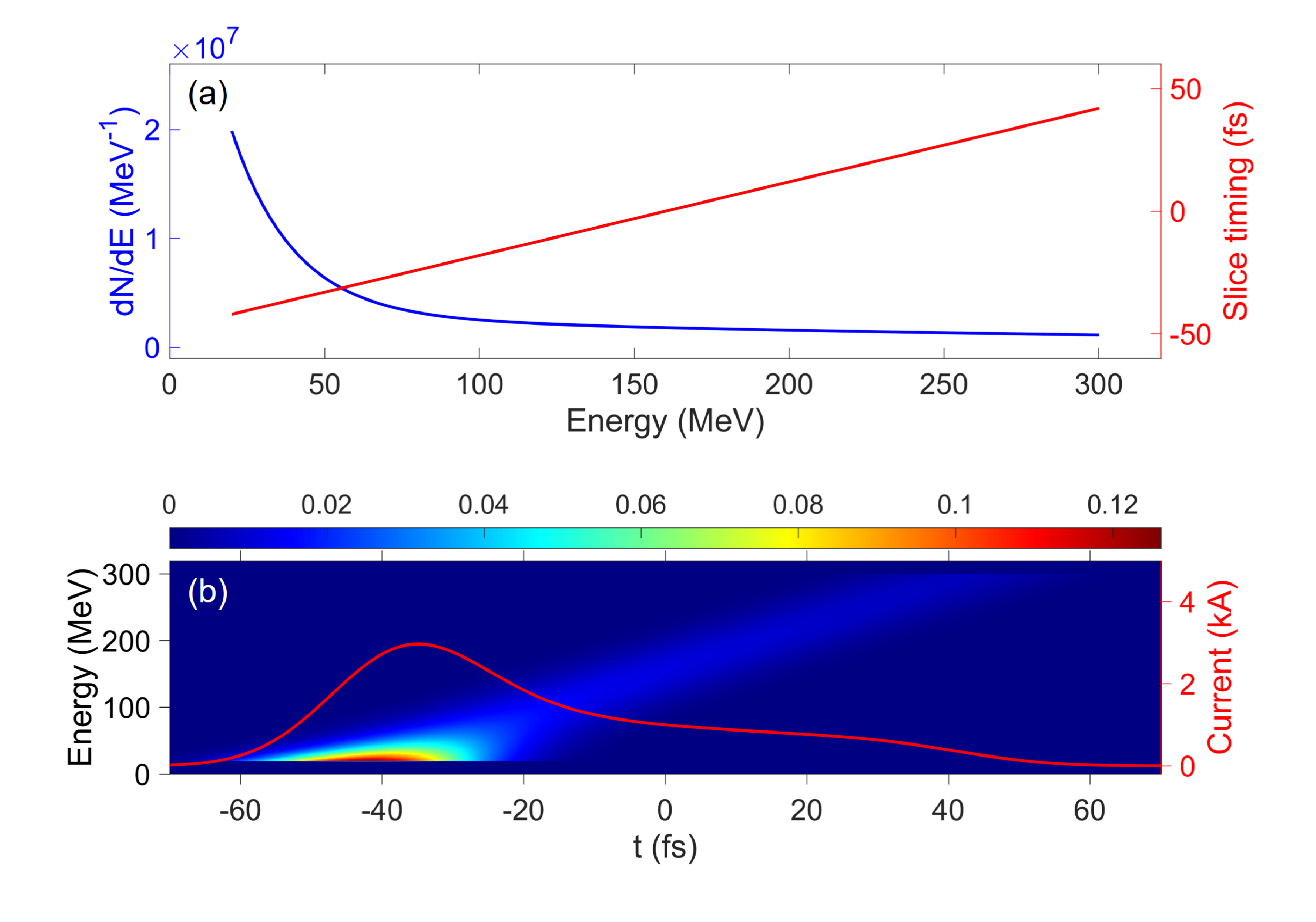}
		\caption{Longitudinal energy chirp and current profiles. (a) The blue line illustrates the two-temperature energy distribution. The red line shows the slice timings of a positive chirp of 0.2 fs/MeV. (b) shows the 2D plot of the LPS of the two-temperature linear-chirped electron bunch. The red curve shows the current profile.}
	\end{figure}
	\begin{figure}[ht]
		\centering
		\includegraphics[width=8.6 cm]{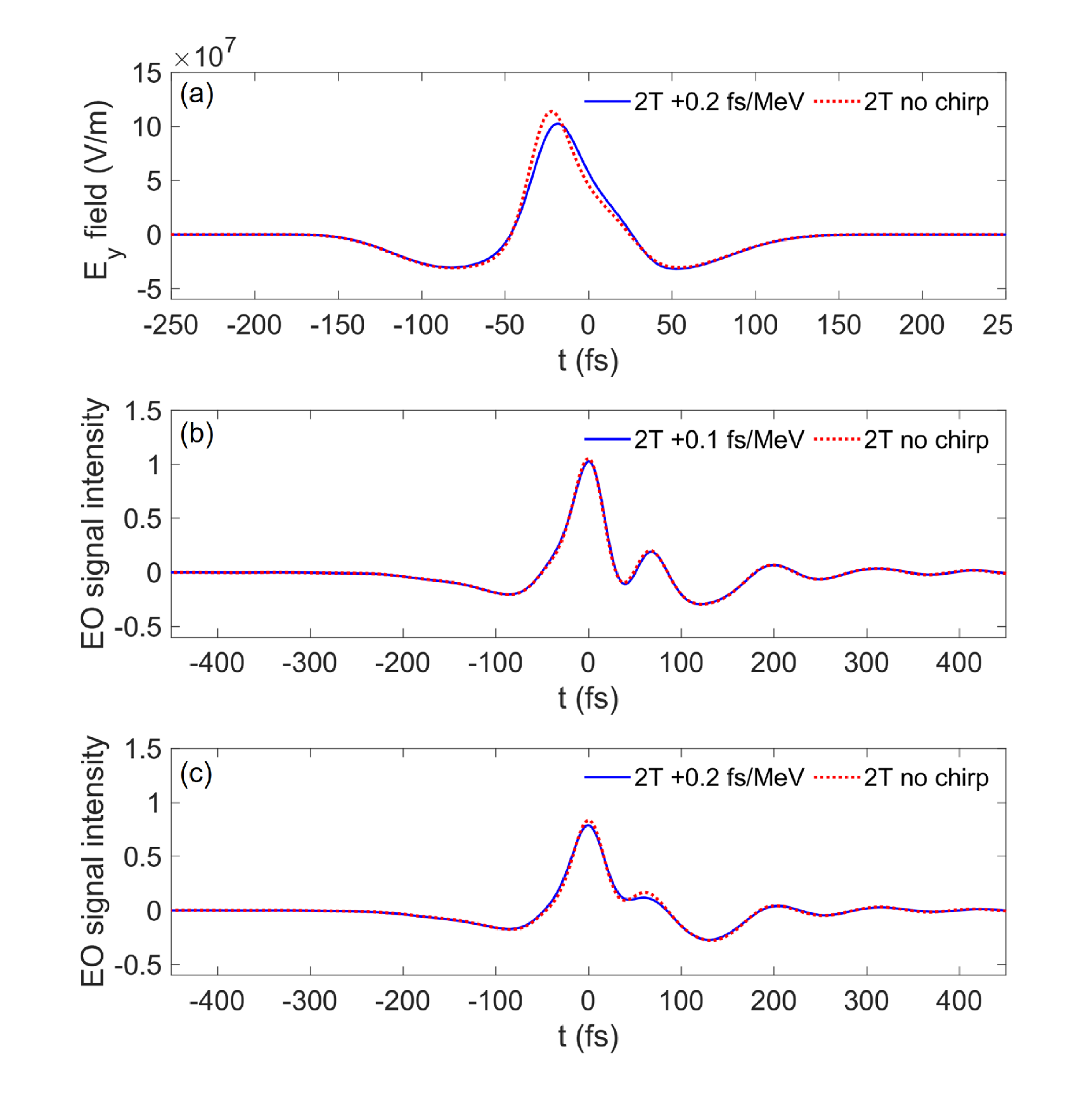}
		\caption{Impact of temporal energy chirp on the EO signals. The blue and red curves illustrate the results of chirped and nonchirped cases, respectively. (a) shows the field strength profiles at ($X_D$, $Y_D$) = (0, -200) $\mu$m. (b) and (c) show the EO signals from current profiles with a chirp of 0.1 and 0.2 fs/MeV, respectively.}
	\end{figure}	
	
	To verify the impact of LPS on the EO signal, we compared the chirped and non-chirped electron bunches with a same current profile and overall charge. The normalized longitudinal profile was retrieved from the current profile, as shown in Fig. 17(b). Subsequently, we calculated the longitudinal form factor $F_z(\omega)$ by performing Fourier transformation. In the non-chirped case, $F_z(\omega)$ was identical for all electron energy components. The field strengths $E_y$ are plotted in Fig. 18(a). The calculated EO signals from the 30-$\mu$m GaP crystal are shown in Figs. 18 (b) and 18(c). Chirps of 0.1 and 0.2 fs/MeV corresponded to temporal spans of 28 and 56 fs, respectively. The field from the chirped electron bunch exhibited a gentle rising edge because the high-energy electrons resided at subsequent timings. However, this small difference did not affect the shape of the EO signals. As shown in Figs. 18(b--c), despite the slight differences in the amplitudes, the shapes of the EO signals were almost identical. \par
	
	These calculations demonstrate that the current profile plays a more important role than the temporal energy chirp in the shapes of the EO signals. When the LPS is unclear, a non-chirped electron bunch is assumed to be sufficient for calculating the longitudinal current distribution. \par

	\section{Noise introduced by the TR when the electrons hit OAP1}
	\begin{figure}[ht]
		\centering
		\includegraphics[width=7.5 cm]{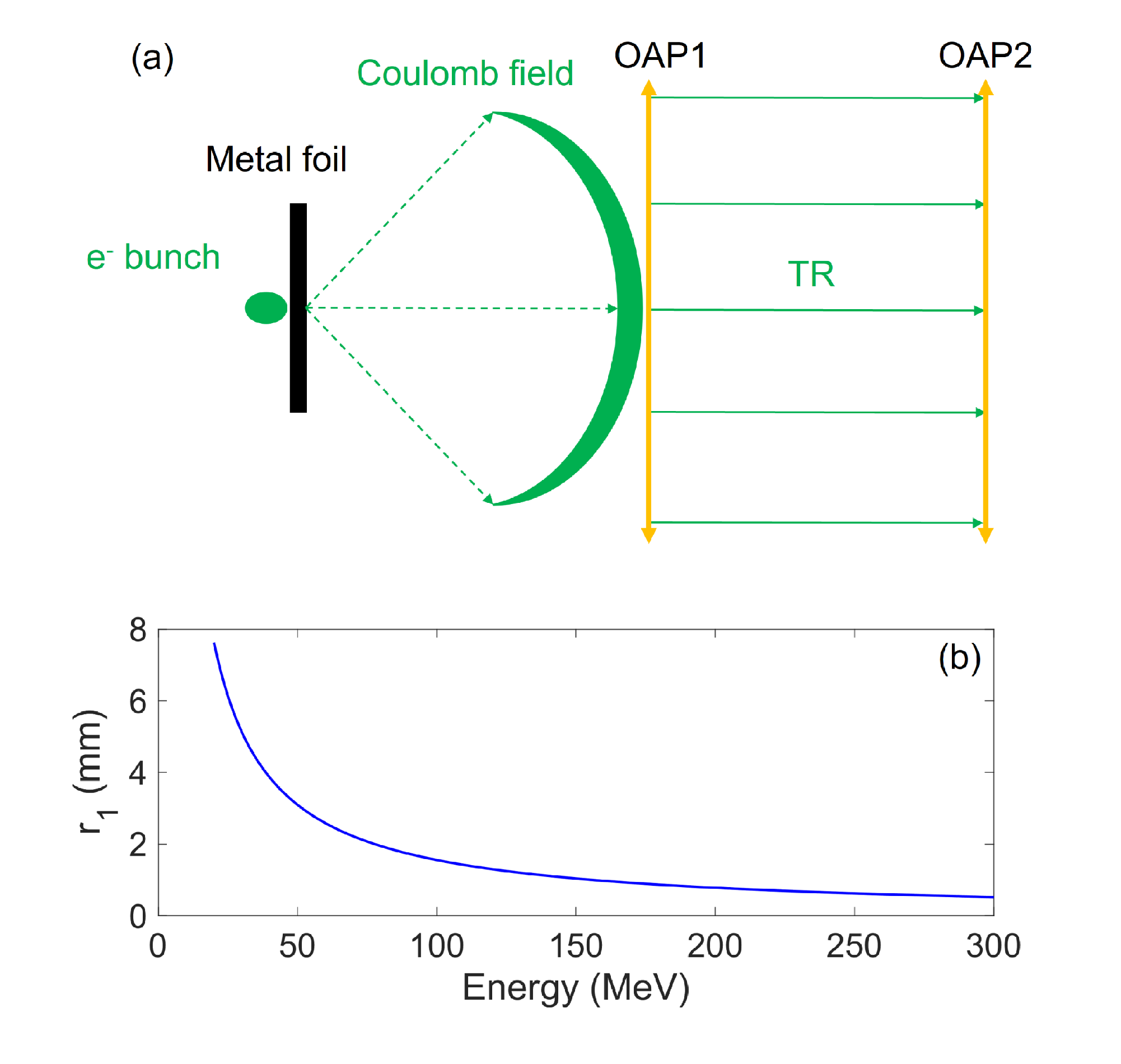}
		\caption{(a) Calculation geometry of the TR from OAP1. TR generated by OAP1 has a plane wave front after OAP1 and is transported to OAP2 with diffraction. (b) Beam size of the scattered electrons at the position of OAP1.}
	\end{figure}
	
	\begin{figure}[ht]
		\centering
		\includegraphics[width=8.8 cm]{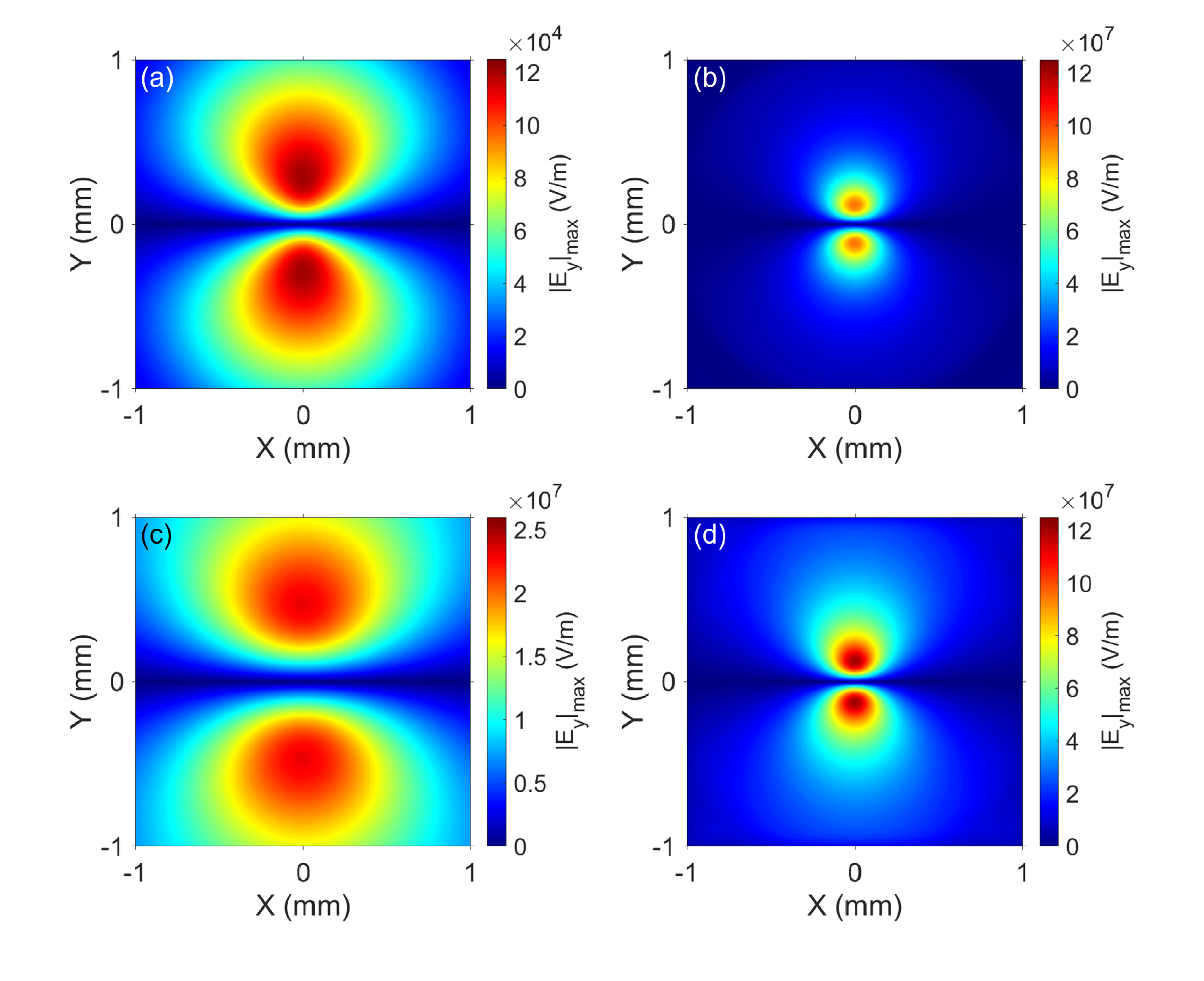}
		\caption{TR field strengths at the EO crystal. (a) TR generated from OAP1 with electron energy of 20 MeV. (b) TR generated from metal foil with electron energy of 20 MeV. (c) TR generated from OAP1 with electron energy of 200 MeV. (d) TR generated from metal foil with electron energy of 200 MeV.}
	\end{figure}
	In the experimental setup illustrated in Fig. 1, electrons passing through the metal foil hit the gold surface of OAP1, causing secondary TR noise. Thus, to remove this noise, the insertion of a bending magnet between the metal foil and OAP1 is recommended. Occasionally, to achieve better spatial resolution, the focal length of OAP1 is set as short, which renders the insertion of a magnet difficult. Here, we evaluated the strength of the TR field produced by OAP1. Figure 19(a) illustrates the calculation process. The scattering caused by the metal foil was calculated using the following equation that is dependent on the composition of the metal, thickness, and electron energy \cite{scattering}:
	\begin{equation}
		\theta_0=\frac{13.6 MeV}{\beta c p}z\sqrt{x/X_0}[1+0.038 \ln(x/X_0)]
	\end{equation}
	\noindent
	where $\theta_0$ is the $rms$ divergence of the scattered beam, $\beta c$ and $p$ are the velocity and momentum of the electrons, respectively, $X_0$ denotes the radiation length of the material, and $x$ denotes the thickness of the material. With a propagation distance of $F$ to OAP1, the electron beam size was $r_1=\sqrt{r_0^2+(\theta_0F)^2}$, where $r_0$ is the beam size at the position of the metal foil. For alloys such as stainless steel, the overall radiation length can be estimated as $1/X_0=\sum_{j}w_j/X_j$, where the $j$th element has a weight of $w_j$ and a radiation length of $X_j$. The radiation length of each element can be found in the NIST database \cite{NISTdatabase}. The weights of the elements of a material are also obtained from various databases. For example, we selected SUS304 stainless steel as the material for the metal foil \cite{SUS304} with a thickness of 100 $\mu$m, an initial beam size of $r_0$ = 50 $\mu$m, and $F$ = 190 mm. The $r_1$ for various electron energies at OAP1 are plotted in Fig. 19(b). The size of the electron beams increased to several millimeters when the electrons had lower energies. For electrons with energies of 20 and 200 MeV, the beam sizes at OAP1 were 7623 $\mu$m and 780 $\mu$m, respectively, resulting in decreased charge densities.  \par
	The electron bunch on the metal foil was considered as the point source. The scattered electrons traversed OAP1 with phases forming a spherical wavefront. Thus, the Coulomb field of the electrons also exhibited a spherical phase. The phase difference on a plane perpendicular to the direction of propagation was described as $\Delta \phi=k(\sqrt{F^2+\rho^2}-F)\approx k \rho^2/2F$. After reflection from OAP1, the spherical phase was corrected using the transform function in Eq.~(\ref{lenstransform}). The TR emitted from OAP1 was calculated as follows: 
	\begin{equation}
		E^{lr1}_{OAP1}=(g_{\perp}^{OAP1}\ast E^S)\times P_l
	\end{equation}
	
	\noindent
	where $g_{\perp}^{OAP1}$ is the transverse distribution of the scattered electron beam at OAP1 and $P_l$ is the pupil defined by the OAP. $E^S$ was calculated using Eq. (1). The procedure described in Section III.A was followed for calculating lateral Huygens--Fresnel diffractions in the optical system. For a fair comparison, we calculated the TR fields at the position of the EO crystal originating from two sources: the metal foil and OAP1. The TR generated from OAP1 with electron energies of 20 and 200 MeV are shown in Figs. 20(a) and 20(c), respectively. The corresponding results of the TR generated from the metal foil are shown in Figs. 20(b) and 20(d), respectively. With an electron energy of 20 MeV, the TR field generated from OAP1 was three orders of magnitude smaller than that generated from the metal foil. Whereas, at a higher electron energy of 200 MeV, the TR field generated from OAP1 was six times smaller than that generated from the metal foil. \par
	
	Apparently, a smaller energy corresponds to a larger scattering angle. This results in a smaller charge density at the position of OAP1. If most of the electron energy is concentrated at approximately 100 MeV or a few tens of MeV, the noise TR source from OAP1 can be ignored. In the experiment, by measuring the electron energy spectrum, the impact of such noise was examined by performing this calculation, including the contributions of all energy components. Notably, temporal elongation of the electron bunches owing to energy spread was not considered. Thus, the noise should be even weaker when calculating the bunch elongation. \par
	
	\section{Discussion and Summary}
	In this study, the shapes of the signals in the EO spatial decoding setup were calculated at arbitrary positions in the image plane. We found that the duration and oscillation behavior of the EO signals varied with parameters such as the electron energy, bunch duration, and transverse beam size. Even with the same electron bunch parameters, the EO signal was broadened further from the center of the TR field. This behavior is similar to cases wherein Coulomb fields were measured \cite{wang2017temporal}. Thus, we did not intend to provide a quick one-to-one correspondence between the duration of the main peak of the EO signal and the original electron bunch duration. Using the methods described herein, such correlations can be identified with specific electron bunch parameters. Spatially resolved detection is strongly recommended when conducting temporal or spectral measurements of a TR field using an imaging system. Furthermore, it is necessary to determine the position of the spatial point at which the experimental signals are measured. \par
	
	In summary, we conducted a systematic numerical study of EO spatial decoding of TR from relativistic electron bunches. 3D TR field was determined using both a detailed calculation based on the Huygens--Fresnel principle and a simplified analytical model based on Fraunhofer assumption. The results suggest that the simplified analytical model is sufficient to perform such polychromatic calculations with considerable accuracy. For EO spatial decoding, we discussed the process of EO signal generation using both 1D and 2D models. The 1D model was concluded to be sufficient for quick data analysis. In addition, we demonstrated the minimal impact of the temporal energy chirp on the shape of the EO signal. Further, we proposed a method to estimate the noise level for unwanted TRs created by reflection optics. Therefore, this study will be useful for investigating the 3D charge-density profiles of ultrafast electron bunches in both LWFA and conventional accelerators. \par

	\begin{acknowledgments}
		We are grateful for the encouragement provided by Dr. Y. Sano, and Dr. N. Kumagai. We thank Dr. J. K. Koga for providing valuable insights into the current research topic. This work was funded by the JST-Mirai Program, Japan (Grant No. JPMJMI17A1); the Grant-in-Aid for Early-Career Scientists (Grant Nos. JP21K17998 and JP23K17152); Grant-in-Aid for Scientific Research (C) from JSPS KAKENHI, Japan (Grant No. JP22K12665); and the QST President's Strategic Grant (Exploratory Research), Japan. \par
	\end{acknowledgments}
	
	\appendix
	\section{Formulation of the electric field of transition radiation in SI units}
	
	Herein, the self-field expression for a single electron is derived. Similar derivations can be found in several textbooks and literature on Gaussian units \cite{jackson1999classical,kube2008imaging,Mikaelian}. We rewrote the formalization in SI units. Maxwell's equations in vacuum are as follows:
	
	\begin{equation} \label{Maxwell}
		\begin{aligned}
			\nabla \cdot \bm{E} &=\dfrac{\rho}{\epsilon_0} \\
			\nabla \cdot \bm{B} &=0 \\
			\nabla \times \bm{E} &=-\dfrac{\partial \bm{B}}{\partial t} \\
			\nabla \times \bm{B} &=\mu_0(\bm{J}+\epsilon_0\dfrac{\partial \bm{E}}{\partial t})
		\end{aligned}
	\end{equation}
	\noindent
	The scalar potential $\varphi$ and vector potential $\bm{A}$ are expressed as,
	\begin{equation} \label{vectorscalar}
		\begin{aligned}
			\bm{B}&=\nabla\times \bm{A} \\
			\bm{E}&=-\nabla \varphi-\dfrac{\partial \bm{A}}{\partial t}
		\end{aligned}
	\end{equation}
	\noindent
	The Lorentz Gauge equation is expressed as,
	\begin{equation} \label{lorentz}
		\nabla\cdot\bm{A}+\dfrac{1}{c^2}\dfrac{\partial \varphi}{\partial t}=0
	\end{equation}
	
	\noindent
	Via Eq.~(\ref{Maxwell}) and Eq.(\ref{vectorscalar}) and Eq.~(\ref{lorentz}), with $c=1/\sqrt{\epsilon_0\mu_0}$, the wave equations can be derived as:
	\begin{equation} \label{waveequations}
		\begin{aligned}
			(\nabla^2-\dfrac{1}{c^2}\dfrac{\partial^2}{\partial t^2})\varphi(\bm{r}, t)&=-\dfrac{\rho(\bm{r}, t)}{\epsilon_0} \\
			(\nabla^2-\dfrac{1}{c^2}\dfrac{\partial^2}{\partial t^2})\bm{A}(\bm{r}, t)&=-\mu_0\bm{J}(\bm{r}, t)
		\end{aligned}
	\end{equation}
	
	\noindent
	Here, the variables ($\varphi$, $\bm{A}$, $\rho$, and $\bm{J}$) are expressed in space and time coordinates ($\bm{r}$ and $t$), where the instant charge, current, and displacement of a single electron are expressed as follows:
	\begin{equation} \label{chargecurrent}
		\begin{aligned}
			\rho(\bm{r},t)&=-e\delta(\bm{r}-\bm{r}(t)) \\
			\bm{J}(\bm{r},t)&=-e\bm{v}\delta(\bm{r}-\bm{r}(t)) \\
			\bm{r}(t)&=\bm{v}t
		\end{aligned}
	\end{equation}
	
	To avoid discrepancies caused by different conventions of Fourier transformation in the angular frequency, we used the standard Fourier transformation as follows:
	
	\begin{equation} \label{fourier4}
		\begin{aligned}
			g_1(\bm{r},t)&=\int{df}\int{d^3\bm{\tilde{\nu}}}g_2(\bm{\tilde{\nu}},f)\exp[i2\pi(\bm{\tilde{\nu}}\cdot\bm{r}-f t)] \\
			g_2(\bm{\tilde{\nu}},f)&=\int{dt}\int{d^3\bm{r}}g_1(\bm{r},t)\exp[-i2\pi(\bm{\tilde{\nu}}\cdot\bm{r}-f t)]
		\end{aligned}
	\end{equation}

	\noindent
	where ($\bm{\tilde{\nu}}$, $f$) and ($\bm{r}$, $t$) are the reciprocal $\bm{\tilde{\nu}}=\bm{k}/2\pi$ and $f=\omega/2\pi$. By performing Fourier transformations on both sides of Eq.~(\ref{waveequations}), using the relationships in Eq.~(\ref{fourier4}), the wave equations in the frequency domain are derived as,
	\begin{equation}
		\begin{aligned}
			(\bm{\tilde{\nu}}^2-\dfrac{f^2}{c^2})\varphi(\bm{\tilde{\nu}},f)&=\dfrac{\rho(\bm{\tilde{\nu}},f)}{4\pi^2\epsilon_0} \\
			(\bm{\tilde{\nu}}^2-\dfrac{f^2}{c^2})\bm{A}(\bm{\tilde{\nu}},f)&=\dfrac{\bm{J}(\bm{\tilde{\nu}},f)}{4\pi^2\epsilon_0 c^2}
		\end{aligned}	
	\end{equation} 
	\noindent
	where $\rho(\bm{\tilde{\nu}},f)$ and $\bm{J}(\bm{\tilde{\nu}},f)$ are derived using Eqs. ~(\ref{chargecurrent}) and ~(\ref{fourier4}) as
	
	\begin{equation}
		\begin{aligned}
			\rho(\bm{\tilde{\nu}},f)&=-e\delta(f-\bm{\tilde{\nu}}\cdot \bm{v}) \\
			\bm{J}(\bm{\tilde{\nu}},f)&=-e\bm{v}\delta(f-\bm{\tilde{\nu}}\cdot \bm{v})
		\end{aligned} 
	\end{equation}
	
	The scalar and vector potentials of the field in the $(\bm{k},\omega)$ domain are obtained as follows:
	
	\begin{equation}
		\begin{aligned}
			\varphi(\bm{\tilde{\nu}},f)&=-\dfrac{e}{(2\pi)^2\epsilon_0}\dfrac{\delta(f-\bm{\tilde{\nu}}\cdot \bm{v})}{\bm{\tilde{\nu}}^2-f^2/c^2} \\
			\bm{A}(\bm{\tilde{\nu}},f)&=-\dfrac{e\bm{v}}{(2\pi)^2\epsilon_0 c^2}\dfrac{\delta(f-\bm{\tilde{\nu}}\cdot \bm{v})}{\bm{\tilde{\nu}}^2-f^2/c^2}
		\end{aligned} 
	\end{equation}
	
	\noindent
	Using the relationships in Eq.(\ref{vectorscalar}), we obtain
	
	\begin{equation}\label{Ekw0}
		\bm{E}(\bm{\tilde{\nu}},f)=i2\pi f\bm{A}(\bm{\tilde{\nu}},f)-i2\pi\bm{\tilde{\nu}}\varphi(\bm{\tilde{\nu}},f)
	\end{equation}
	
	\noindent
	The formula for the electric field $E(\bm{k},\omega)$ is derived as follows:
	\begin{equation}\label{Ekw1}
		\bm{E}(\bm{\tilde{\nu}},f)=-i\dfrac{e}{2\pi\epsilon_0}\dfrac{f\bm{v}/c^2-\bm{\tilde{\nu}}}{\bm{\tilde{\nu}}^2-f^2/c^2}\delta(f-\bm{\tilde{\nu}}\cdot \bm{v})
	\end{equation}
	
	For simplicity, we considered electron propagation along the z-axis. We have $\bm{v}=(0,0,v)$ and $f=\bm{\tilde{\nu}}\cdot \bm{v}=\tilde{\nu}_zv$, where $\bm{\tilde{\nu}}=(\tilde{\nu}_x,\tilde{\nu}_y,\tilde{\nu}_z)$. With the property of the delta function $\delta(ax)=\delta(x)/|a|$, Eq.(\ref{Ekw1}) is rewritten as follows:
	
	\begin{equation}\label{Ekw2}
		\bm{E}(\bm{\tilde{\nu}},f)=i\dfrac{e}{2\pi\epsilon_0 v}\delta(f/v-\tilde{\nu}_z)\dfrac{(\tilde{\nu}_x,\tilde{\nu}_y,\tilde{\nu}_z/\gamma^2)}{\tilde{\nu}_x^2+\tilde{\nu}_y^2+\tilde{\nu}_z^2/\gamma^2} 
	\end{equation}
	
	Thus, the frequency-domain electric field in space can be obtained using the integral in Eq.~(\ref{Erw1}), with $\bm{r}=(x,y,z)$. As the longitudinal component of the field was weaker than the transverse component by a factor of $1/\gamma^2$, we focused on analyzing the transverse component of the electric field in this study. The $(E_x,E_y)$ components were calculated using Eq. (~\ref{Erwxy11}).
	
	\begin{widetext}
		\begin{eqnarray}\label{Erw1}
			\bm{E}(\bm{r},f)=i\dfrac{e}{2\pi\epsilon_0 v}\int{d^3\bm{\tilde{\nu}}}\exp[i2\pi(\tilde{\nu}_xx+\tilde{\nu}_yy+\tilde{\nu}_zz)]\delta(f/v-\tilde{\nu}_z)\dfrac{(\tilde{\nu}_x,\tilde{\nu}_y,\tilde{\nu}_z/\gamma^2)}{\tilde{\nu}_x^2+\tilde{\nu}_y^2+\tilde{\nu}_z^2/\gamma^2} 
		\end{eqnarray}
		
		\begin{eqnarray}\label{Erwxy11}
			E_{x,y}(\bm{r},f)=i\dfrac{e}{2\pi\epsilon_0 v}\exp(i2\pi f z/v)\iint{d\tilde{\nu}_xd\tilde{\nu}_y}\exp[i2\pi(\tilde{\nu}_xx+\tilde{\nu}_yy)]\dfrac{(\tilde{\nu}_x,\tilde{\nu}_y)}{\tilde{\nu}_x^2+\tilde{\nu}_y^2+f^2/v^2\gamma^2}
		\end{eqnarray}
		
		\begin{eqnarray}\label{Erwxy12}
			E_{x,y}(\bm{r},f)=-\frac{ef/\gamma v}{\epsilon_0 v}\exp(i2\pi f z/v)\frac{(x,y)}{\sqrt{x^2+y^2}}K_1(2\pi f/\gamma v \sqrt{x^2+y^2})
		\end{eqnarray}
		
		\begin{eqnarray}\label{Erwxy21}
			E_{x,y}(\bm{r},\omega)=i\dfrac{e}{(2\pi)^2\epsilon_0 v}\exp(i\omega z/v)\iint{dk_xdk_y}\exp[i(k_xx+k_yy)]\dfrac{(k_x,k_y)}{k_x^2+k_y^2+\alpha^2}
		\end{eqnarray}
		\begin{eqnarray}\label{Erwxy22}
			E_{x,y}(\bm{r},\omega)=-\frac{e\alpha}{2\pi\epsilon_0 v}\exp(i\omega z/v)\frac{(x,y)}{\sqrt{x^2+y^2}}K_1(\alpha \sqrt{x^2+y^2})
		\end{eqnarray}
	\end{widetext}
	
	By calculating the integral in Eq. ~(\ref{Erwxy11}), the transverse field strength is expressed as Eq.~(\ref{Erwxy12}), where $K_1$ is a Bessel function of the second kind. By introducing a parameter $\alpha=\omega/\gamma v=2\pi f/\gamma v$, Eqs.~(\ref{Erwxy11}) and ~(\ref{Erwxy12}) were changed to the form described by the angular frequency in Eqs.~(\ref{Erwxy21}) and ~(\ref{Erwxy22}). Notably, the factor $\dfrac{e\alpha}{2\pi\epsilon_0 v}$ in Eqs. ~(\ref{Erwxy22}) differs from $\dfrac{e\alpha}{(2\pi)^{3/2}\epsilon_0 v}$ in \cite{casalbuoni2009ultrabroadband,brau2004modern}, where the unitary convention of Fourier transformation $g(t)=\frac{1}{(2\pi)^{1/2}}\int g(\omega)\exp(-i\omega t)dt$ was applied. \par	
	
	The derivation of Eq.~(\ref{Erwxy12}) required the integrals \cite{integraltable} listed as follows: \par
	
	\begin{subequations} \label{specialintegral}
		
		\begin{equation}
			\int_{-\infty}^{+\infty}\dfrac{x\sin{ax}}{b^2+x^2}dx=\pi e^{-ab}
		\end{equation}
		\begin{equation}
			\int_{0}^{+\infty}e^{-a\sqrt{c^2+x^2}}\cos(bx)dx=\dfrac{ac}{\sqrt{a^2+b^2}}K_1(c\sqrt{a^2+b^2})
		\end{equation}
	\end{subequations}

	\begin{figure}[ht]
		\centering
		\includegraphics[width=8.6 cm]{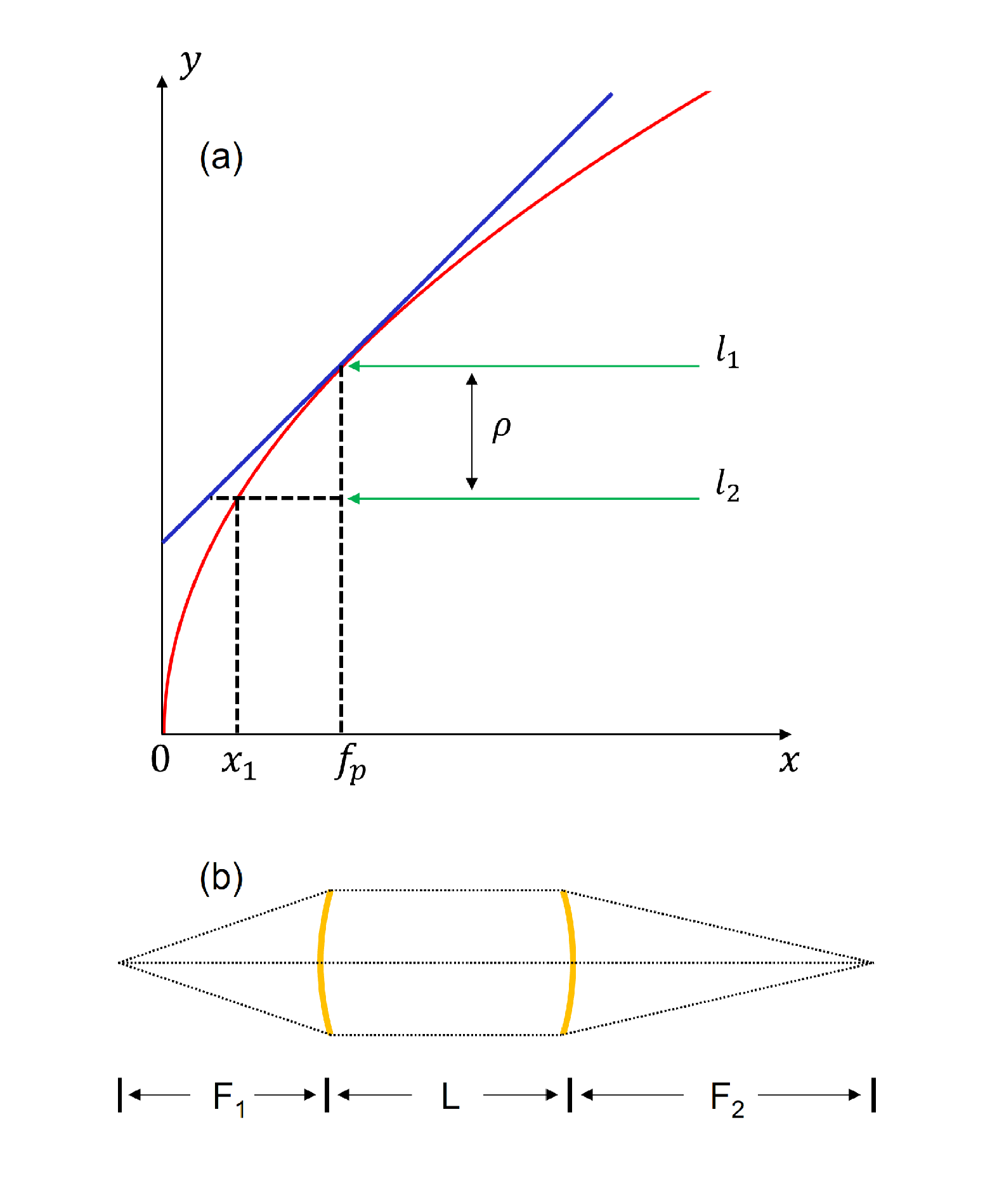}
		\caption{(a) Explanation of the phase shift introduced by a parabolic mirror. (b) The overall transmission process from the source to the detector in a 2-OAP imaging system.}
		\label{parabola}
	\end{figure}
	
	\section{Imaging system with two OAPs}

	Herein, a simple explanation of the imaging system composed of two OAPs is presented. We observed that, geometrically, such a system can be considered as a thin lens with additional phase delay. The phase-shift effect of the parabolic mirror is shown in Fig. 21(a). The parabolic surface, plotted as a red curve, is defined by the function $y^2=4f_px$, where $f_p$ is the parent focal length of the parabola. The light indicated by $l_1$ was incident at the center of a 90$\degree$ parabolic mirror. The sideline $l_2$ has a transverse distance $\rho$ to $\l_1$. The tangent of the parabola at point $(f_p, 2f_p)$ has a slope of 1 and a relative angle of 45$\degree$ on the x-axis. Without a parabolic mirror, the light was reflected by a flat mirror, as illustrated by the blue line, and the overall optical paths were the same. The red curve indicates that $l_2$ is incident at point $(x_1, 2f_p-\rho)$, where $x_1=(2f_p-\rho)^2/4f_p$. Thus, the introduced path difference is $f_p-x_1-\rho$, which is calculated as follows: 
	
	\begin{equation}
		\Delta L(\rho)=-\dfrac{\rho^2}{4f_p}
	\end{equation}
	
	\noindent
	For the 90$\degree$ OAP, the effective focal length was $f_e=2f_p$. For the 2-OAP system shown in Fig. 21(b), the overall transformation function is calculated as $T(\rho)=\exp[ik(F_1+L+F_2)]\exp[ik(\Delta L_1(\rho)+\Delta L_2(\rho))]$. Neglecting the constant phase delay, the transformation can be written as
	
	\begin{equation}
		\begin{aligned}
			T(\rho)&=\exp[-\dfrac{ik\rho^2}{2}(\dfrac{1}{F_1}+\dfrac{1}{F_2})] \\
			&=\exp[-\dfrac{ik\rho^2}{2}\dfrac{1}{f}]
		\end{aligned}
	\end{equation}
	
	\noindent
	where $1/f=1/F_1+1/F_2$ and $f$ is the effective focal length of the overall system.  By placing the source at the focal point of the first OAP, this imaging system had an effective focal length of $f$ and a magnification of $F_2/F_1$. This calculation was performed based on geometrical optics. \par

	\section{Derivation of the formula for TR imaging using Fraunhofer approximation}
	The derivation of an analytical formula in the frequency domain for OTR imaging has been reported in existing literature \cite{xiang2007theoretical,CastellanoOTR}. Here, we listed the details used to confirm the factors in SI units. The imaging system comprised only one thin lens. The propagation of TR was separated into two sessions: (i) source to lens and (ii) lens to detector. Assuming a metal foil of infinite size and neglecting the spherical phase in both the source and detector planes, the overall diffractive propagation can be expressed as Eq.~(\ref{Fraunhoferimagingxy}) using the Fraunhofer approximation. \par
	
	\begin{widetext} 
		
		\begin{multline}\label{Fraunhoferimagingxy}
			E^D(X_D,Y_D)=\frac{\exp(ikF_1)}{i\lambda F_1}\frac{\exp(ikF_2)}{i\lambda F_2}\int_{X_{l,min}}^{X_{l,max}}dX_l\int_{Y_{l,min}}^{Y_{l,max}}dY_l  \exp[-\frac{ik}{F_2}(X_DX_l+Y_DY_l)]   \\
			\times \exp[-\frac{ik}{2F_2}(X_l^2+Y_l^2)]\exp[-\frac{ik}{2F_1}(X_l^2+Y_l^2)]\exp[\frac{ik}{2F_2}(X_l^2+Y_l^2)]\exp[\frac{ik}{2F_1}(X_l^2+Y_l^2)] \\
			\times \int_{-\infty}^{+\infty}dX_s\int_{-\infty}^{+\infty}dY_s E^s(X_s,Y_s)\exp[-\frac{ik}{F_1}(X_sX_l+Y_sY_l)]
		\end{multline}

	\end{widetext}

	\begin{widetext} 
		
		\begin{multline}\label{Fraunhoferimagingr}
			E^D(\rho_D,\phi_D)=\frac{e\alpha}{2\pi\epsilon_0v}\frac{1}{\lambda^2M}\int_{0}^{\theta_m}\theta d\theta\int_{0}^{2\pi}d\phi_l\exp[-\frac{ik}{M}\rho_D\theta\cos(\phi_D-\phi_l)] \\
			\times\int_{0}^{+\infty}K_1(\alpha\rho_s)\rho_s d\rho_s \int_{0}^{2\pi}d\phi_s(\cos\phi_s,\sin \phi_s)\exp[-ik\rho_s\theta\cos(\phi_s-\phi_l)]
		\end{multline}

	\end{widetext}
	

	This system exhibits cylindrical symmetry because the pupil defined by the OAP has a circular shape. We change the notations of the coordinates as: $(X_s, Y_s)=\rho_s(\cos\phi_s,\sin\phi_s)$, $(X_l, Y_l)=\rho_l(\cos\phi_l,\sin\phi_l)$, $(X_D, Y_D)=\rho_D(\cos\phi_D,\sin\phi_D)$. The Eq.~(\ref{Erwxy22}) is rewritten as $E^s(\rho_s)=-\frac{e\alpha}{2\pi\epsilon_0 v}\exp(i\omega z/v)K_1(\alpha\rho_s)$.  With  $\theta=\rho_l/F_1$ and $M=F_2/F_1$, Eq.(\ref{Fraunhoferimagingxy}) can be simplified as Eq.~(\ref{Fraunhoferimagingr}) by omitting all longitudinal phase factors. $\theta_m=R_l/F_1$ denotes the acceptance angle of the first OAP. Using the integral $\int_{0}^{2\pi}d\beta_1(\cos\beta_1,\sin\beta_1)\exp[-iA\cos(\beta_1-\beta_2)]=-2\pi i (\cos\beta_2,\sin\beta_2)J_1(A)$ and the definition of the normalized transverse distance in the image plane $\zeta=k\rho_D/M$, Eq.~(\ref{Fraunhoferimagingr}) can be simplified as
	\begin{multline}
		E^D(\rho_D,\phi_D)=\frac{-ek\alpha}{\epsilon_0 v \lambda M}(\cos\phi_D,\sin \phi_D)\int_{0}^{\theta_m}\theta J_1(\zeta \theta) \\
		\times \int_{0}^{+\infty}\rho_s d\rho_s J_1(k\theta \rho_s)K_1(\alpha \rho_s)
	\end{multline}
	
	By using the integral $\int_{0}^{+\infty}xJ_1(ax)K_1(bx)dx=a/(a^2b+b^3)$ \cite{integraltable}, we have
	\begin{multline}\label{rhofinal}
		E^D(\rho_D,\phi_D)=-\frac{e}{\epsilon_0}\frac{(\cos\phi_D,\sin \phi_D)}{\lambda M v} \\
		\times \int_{0}^{\theta_m}\frac{\theta^2}{\theta^2+(\gamma\beta)^{-2}}J_1(\zeta\theta)d\theta
	\end{multline}
	
	In most cases, $\beta \simeq 1$. Equation(\ref{rhofinal}) can be expressed in Cartesian coordinates as follows:
	
	\begin{multline}\label{xyfinal}
		E^D(X_D,Y_D)=-\frac{e}{\epsilon_0}\frac{1}{\lambda M v}\frac{(X_D, Y_D)}{\sqrt{X_D^2+Y_D^2}} \\
		\times \int_{0}^{\theta_m}\frac{\theta^2}{\theta^2+\gamma^{-2}}J_1(\zeta\theta)d\theta
	\end{multline}

\end{document}